\documentclass[%
 reprint,
 amsmath,amssymb,
 aps,
floatfix,
]{revtex4-1}

\usepackage{subcaption}
\usepackage{standalone}
\usepackage{comment}
\usepackage{changepage}

\usepackage{float}

\usepackage{graphicx}
\usepackage{dcolumn}
\usepackage{bm}

\chardef\MyArticleWithColor=\pdfcolorstackinit page direct{0 g}

\DeclareMathOperator{\sgn}{sgn}

\usepackage{tikz}
\usepackage{tikz-feynman}
\tikzfeynmanset{warn luatex=false}
\usepackage{cellspace}
\usepackage{adjustbox}

\newcommand*{\halfway}{0.5*\pgfdecoratedpathlength+.5*7pt}
\tikzset{->-/.style={decoration={markings, mark=at position #1 with {\arrow{latex}}},postaction={decorate}},
	->-/.default=\halfway}

\begin{document}


\title{Loop Tree Duality for multi-loop numerical integration}

\author{Zeno Capatti}%
\email{zeno.ca@gmail.com}
\author{Valentin Hirschi}%
\email{valentin.hirschi@gmail.com}
\author{Dario Kermanschah}%
\email{d.kermanschah@gmail.com}
\author{Ben Ruijl}%
\email{benruyl@gmail.com}
\affiliation{ETH Z\"urich,\\
R\"amistrasse 101, %
8092 Z\"urich, Switzerland}

\date{\today}

\begin{abstract}
Loop Tree Duality (LTD) offers a promising avenue to numerically integrate multi-loop integrals directly in momentum space. It is well-established at one loop, but there have been only sparse numerical results at two loops. We provide a formal derivation for a novel multi-loop LTD expression and study its threshold singularity structure. We apply our findings numerically to a diverse set of up to four-loop finite topologies with kinematics for which no contour deformation is needed. We also lay down the ground work for constructing such a deformation. Our results serve as an important stepping stone towards a generalised and efficient numerical implementation of LTD, applicable to the computation of virtual corrections.
\end{abstract}

\maketitle


\section{\label{sec:introduction}Introduction}

Loop integrals are an essential component of fixed-order corrections to collider cross-sections. 
Analytic techniques enjoyed a durable success in this matter, but it has becomes increasingly evident that a further breakthrough necessitates a radical change of perspective. Numerical approaches are a promising alternative and have already been extensively explored for Feynman amplitudes using sector decomposition (see e.g.~\cite{Hepp:1966eg,Roth:1996pd,Binoth:2000ps,Anastasiou:2005pn,Heinrich:2008si,Smirnov:2015mct,Borowka:2017idc}). More recently, direct integration of finite loop integrals in four-dimensional Minkowskian momentum space have been considered, together with the necessary complex contour deformation for handling integrable threshold singularities~\cite{Gong:2008ww, BeckerMultiLoop2012,BeckerMasses2012,Anastasiou:2008rm}.
In this letter we study the possibility of rewriting an $n-$loop integral as a sum of terms with $n$ additional on-shell conditions by analytically integrating over loop energies using residue theorem. The ensuing identity is called Loop Tree Duality~\cite{LTDRodrigoOrigin2008} (LTD).
LTD is appealing from a numerical standpoint for at least four reasons: (1) the $n-$loop integral dimensionality is fixed to 3$n$ irrespective of the topology considered, (2) integrable singularities can be shown to be confined to a bounded volume~\cite{Aguilera-Verdugo:2019kbz} and are absent when considering certain kinematical configurations, (3) momentum-space divergent integrals naturally lend themselves to be regularised with local UV and IR counterterms~\cite{Becker:2010ng,Becker:2011aa,Becker:2012aqa,Becker:2012nf,Sborlini:2016gbr,Sborlini:2016hat,Seth:2016hmv,Driencourt-Mangin:2017gop,Anastasiou:2018rib,Baumeister:2019rmh} or even (4) through a direct combination with the corresponding real-emission contributions in the case of physical amplitudes~\cite{Hernandez-Pinto:2015ysa,Page:2018ljf,Runkel:2019zbm}.

In this work we derive a novel multi-loop LTD expression by iteratively applying the residue theorem, carefully keeping track of the propagation of Feynman's causal prescription. This differs from the expression derived in ref.~\cite{LTDRodrigoMultiLoop2010} where distributional identities between Feynman and dual propagators are used. It also differs from the work of ref.~\cite{LTDWeinzierl2019} which seeks to achieve a similar goal using a multidimensional version of the residue theorem, arriving at an expression which is incompatible with ours.

We proceed to numerically apply our LTD construction to various scalar loop topologies ranging from one to four loops. In all cases, we find agreement to better than $1\%$, thereby validating our procedure and the expected structure of integrable as well as cancelling singularities exhibited by each term of our LTD expression. We also determine the constraints on a contour deformation and construct deformation vectors satisfying them. A first preliminary result is given for a two-loop LTD integration using a contour deformation.

The outline of this work is as follows. In section~\ref{sec:ltd_formalism} we derive our LTD expression for an arbitrary number of loops.
In section~\ref{sec:numerics} we discuss our numerical implementation and results. In section~\ref{sec:outlook} we present first steps towards a general contour deformation. Finally, we give our conclusions in section~\ref{sec:conclusion}.

\section{\label{sec:ltd_formalism}\label{sec:ltd_formalism_our_result}Loop Tree Duality formalism}

We consider the following general expression for an $n$-loop integral
\begin{equation}
I = \int \prod\limits_{j=1}^n \frac{\mathrm{d}^4k_j}{(2\pi)^4} \frac{N}{\prod\limits_{i\in \mathbf{e}} D_i},
\quad
D_i = q_i^2-m_i^2+\mathrm{i}\delta,
\label{eq:n-loop}
\end{equation}
where $\mathbf{e}$ is the set of indices labelling the edges of a Feynman diagram and the numerator $N$ is a regular function of the loop momenta. The Feynman propagator $1/D_i$ depends on the four-momentum $q_i\equiv (q_i^0, \vec{q}_i)$, the mass $m_i$ and the positive causal prescription $\mathrm{i}\delta$. We consider pairwise distinct Feynman propagators, each with two first order poles in $q^0_i$ located at $\sigma E_i \equiv \sigma \sqrt{\vec{q}_i^{\ 2}+m_i^2 -\mathrm{i}\delta}$, $\sigma \in \{\pm1\}$, where $+E_i$ lies in lower complex half-plane.
Introducing the signature vector $\mathbf{s}_i = (s_{i1},\dots,s_{in})$, $s_{ij} \in \{\pm1,0\}$, we write $q_i^\mu =\sum_{j=1}^n s_{ij} k_j^\mu + p_i^\mu$, where $p_i^\mu$ is a shift that depends on external momenta. \par

The integration over the momenta $k_j$ can be split up in an integration over a spatial part $\vec{k}_j$ and the energy $k_j^0$. We now derive our LTD formula by performing the energy integrations one after the other, following an arbitrary fixed order of the energy variables $\mathbf{k^0} = (k^0_1,\dots, k^0_n)$. We construct this iterative procedure by considering a contour for each energy integration variable along the real line, and closing on an arc in \emph{either} the upper (with winding number $\Gamma_j = +1$) \emph{or} the lower ($\Gamma_j = -1$) complex half-plane. We assume the integral along the arc to vanish, such that the integral along the real line equals the sum of residues at poles located within the contour. The iterative computation of each loop energy integration yields
\begin{align}
I
=
\int \prod\limits_{j=1}^n \frac{\mathrm{d}^{3}\vec{k}_{j}}{(2\pi)^{3}} (\Gamma_j\mathrm{i})
\sum_{\substack{
\mathbf{i} \in \mathcal{I} \\
\boldsymbol \sigma \in \{\pm1\}^n \\
}}
\mathrm{Res}_{\mathbf{k^{\boldsymbol \sigma}_\mathbf{i}}}[f]
\prod_{r=1}^n
\Theta\left(\Gamma_r\Im [k^{\boldsymbol\sigma}_{\mathbf{i},r}]\right),
\label{eq:masterwiththeta}
\end{align}
where we introduce the set of ordered lists of edge indices $\mathcal{I} = \{(i_1,\dots,i_n) \in \mathbf{e}^n\vert \det\left((s_{i_jj})_{1\leq j \leq r}\right) \neq 0,\ \forall r \leq n\}$, which guarantees that for every iteration $j$, where we integrate out $k_j^0$, the propagator labelled by $i_j$ depends on $k_j^0$. Note that this set can contain several permutations of the same indices. The residue of $f=N/\prod_{i\in \mathbf{e}} D_i$ is
\begin{equation}
\label{eq:startingresidue}
\mathrm{Res}_{\mathbf{k^{\boldsymbol \sigma}_\mathbf{i}}}[f] =
\frac{1}{\det \mathbf{s}_\mathbf{i}}
\frac{1}{\prod_{r=1}^n \sigma_r}
\frac{1}{\prod\limits_{i\in \mathbf{i}}2E_{i}}
\frac{N}{\prod\limits_{i \in \mathbf{e} \setminus \mathbf{i}} D_i}
\Bigg\rvert_{\mathbf{k^0} = \mathbf{k}^{\boldsymbol\sigma}_\mathbf{i}},
\end{equation}
being evaluated at the pole locations implicitly defined through the solutions $\mathbf{k^0}=\mathbf{k}^{\boldsymbol\sigma}_\mathbf{i}$to the following linear system
\begin{equation}
\label{system-residues}
\begin{pmatrix} \sigma_{1} E_{i_1}\\ \vdots \\ \sigma_{n} E_{i_n}\end{pmatrix} 
= 
\begin{pmatrix}
& & \mathbf{s}_{i_1} & &\\
&&\vdots && \\
\multicolumn{5}{c}{\mathbf{s}_{i_n}} \\
\end{pmatrix}
\begin{pmatrix} k^0_1 \\ \vdots \\ k^0_n\end{pmatrix}
+ \begin{pmatrix} p^0_{i_1}\\ \vdots \\ p^0_{i_n}\end{pmatrix} 
\equiv \mathbf{s}_{\mathbf{i}} \cdot \mathbf{k^0} + \mathbf{p^0_i},
\end{equation}
where the signature matrix $\mathbf{s_i}$ is a totally unimodular matrix.
Each residue contributes to the integral if the pole location is within all contours of energy integrals already performed, corresponding to the condition $\Gamma_r \Im [k_{\mathbf{i},r}^{\boldsymbol\sigma}] > 0,\ \forall r \leq n$. The imaginary part of the poles in the energy variable $k^0_r$ is computed using Cramer's rule for the last row of the subsystem of \eqref{system-residues} arising after every iteration. Its final expression is given by
\begin{equation}
\Im [k_{\mathbf{i},r}^{\boldsymbol\sigma}]
= \frac{\det
\begin{pmatrix}
 & \sigma_{1} \Im [E_{i_1}] \\
(s_{i_{j_1}j_2})_{\substack{1\leq j_1\leq r \\1\leq j_2 < r}} &  \vdots \\
& \sigma_{r} \Im [E_{i_r}] \\
\end{pmatrix}
}{\det\left((s_{i_jj})_{1\leq j\leq r}\right)},
\label{eq:imaginaryinterplay}
\end{equation}
which explicitly shows how the imaginary parts of poles selected by previous iterations propagate to the imaginary part of the pole contributing at iteration $r$. \par

Eq.~\eqref{eq:masterwiththeta} contains Heaviside functions $\Theta$ with complicated arguments. However,
we have checked that for \emph{all} topologies from one to six loops the Heaviside functions that do not identically evaluate to either $0$ or $1$ cancel pairwise.
In fact, we find that for each loop momentum basis of the corresponding loop graph, only one combination of energy signs contributes to $I$ with a definite prefactor $(-\mathrm{i})^n$. We call this combination of signs the \emph{cut structure}.
Therefore, we conjecture that eq.~\eqref{eq:n-loop} can be written as
\begin{align}\label{eq:masterformula}
I = (-\mathrm{i})^n &\int \prod\limits_{j=1}^n \frac{\mathrm{d}^{3}\vec{k}_j}{(2\pi)^{3}} \sum_{\mathbf{b} \in \mathcal{B}}
\mathrm{Res}_\mathbf{b}[f],
\end{align}
where $\mathbf{b}$ is the set of all edge indices labelling a loop momentum basis and $\mathcal{B}$ is the set of these sets for all loop momentum bases.
The cut structure of $\mathbf{b}$ is denoted with $\boldsymbol\sigma^\mathbf{b}$.
Each loop momentum basis is assigned a residue, henceforth referred to as a \emph{dual integrand}, reading 
\begin{align}
\label{eq:masterresidue}
\mathrm{Res}_\mathbf{b}[f]
&=
\frac{1}{\prod\limits_{i \in \mathbf{b}} 2E_{i}}
\frac{N}{
\prod\limits_{i\in \mathbf{e}\setminus\mathbf{b}} D_i}
\Bigg\rvert_{\{q^0_j = \sigma^\mathbf{b}_j E_j\}_{j\in \mathbf{b}}}\;,
\end{align}
where solving ${\{q^0_j = \sigma^\mathbf{b}_j E_j\}_{j\in \mathbf{b}}}$ yields ${\mathbf{k^0} = \mathbf{k}^{\boldsymbol\sigma^\mathbf{b}}_\mathbf{b}}$.
Note that the dual integrand is invariant under permutations of the elements in $\mathbf{b}$, unlike eq.~\eqref{eq:startingresidue} that depends on the ordering within $\mathbf{i}$.
Furthermore, the complement $\mathbf{t} = \mathbf{e}\setminus \mathbf{b}$ is the \emph{spanning tree} of the graph.
There is a one-to-one correspondence between a spanning tree $\mathbf{t}$ and a loop momentum basis $\mathbf{b}$, hence the name \emph{Loop Tree Duality}. \par

We expect the sum of residues obtained from analytic integration of loop energies to be independent of the specific loop momentum routing as well as choice of contour closure for each loop energy integration. We verified that these expectations are met by explicitly applying eq.~\ref{eq:masterformula} for various choices of routing and contour closures, each time retrieving the same numerical result for the sum of residues $\sum_{\mathbf{b} \in \mathcal{B}}
\mathrm{Res}_\mathbf{b}[f]$ for given numerical inputs $\vec{k}_j$.
In performing these checks, it was convenient to have the cut structure construction algorithm automated and we provide the corresponding {\sc\small Python} implementation as ancillary material.

\subsection{\label{sec:surfaces} Singular surfaces}

Performing the energy integrations introduces additional dependencies on the regulator $\delta$ in the integrand $\sum_{\mathbf{b} \in \mathcal{B}}
\mathrm{Res}_\mathbf{b}[f]$.
For vanishing $\delta$, the \emph{dual propagator} associated with the loop momentum basis~$\mathbf{b}$ reads
\begin{equation}
\frac{1}{D_i\rvert_{\{q^0_j = \sigma^\mathbf{b}_j E_j\}_{j\in \mathbf{b}}}}
=
\frac{1}{(q_i^0\rvert_{\{q^0_j = \sigma^\mathbf{b}_j E_j\}_{j\in \mathbf{b}}})^2 - (E_i)^2}
\end{equation}
and still features singularities if it can go on-shell.
The inverse dual propagator vanishes on two \emph{singular surfaces}
\begin{equation}\label{implicit-equation-surfaces}
S_i^{\mathbf{b},\sigma}: \
\Delta_{i}^{\sigma,\mathbf{b}}\equiv p_i^{0,\mathbf{b}} + \sigma
E_i +
\sum_{j \in \mathbf{b}} s_{ij}^\mathbf{b} \sigma^\mathbf{b}_j E_j = 0,
\end{equation}
where $\sigma \in \{\pm1\}$, and where $s_{ij}^{\mathbf{b}}$ and $p_i^{0,\mathbf{b}}$ are defined implicitly through the change of basis $q_i^0=\sum_{j\in\mathbf{b}} s_{ij}^{\mathbf{b}}q_j^0+p_i^{0,\mathbf{b}}$.

The singular surfaces can be separated into two classes, which we call \emph{E-} and \emph{H-surfaces.}
To distinguish them, we define the \emph{surface signs} for the surface $S_i^{\mathbf{b},\sigma}$ as the list $\mathcal{S}_i^{\mathbf{b},\sigma} = \{ s_{ij}^\mathbf{b}\sigma^\mathbf{b}_j,\ \forall j\in \mathbf{b}\vert s_{ij}^\mathbf{b} \neq 0\}\cup \{\sigma\}$.
A singular surface where all surface signs are equal is called an E-surface, since its defining equation is the one of an ellipsoid when $n-1$ loop momenta are kept fixed.
Otherwise, it is called an H-surface, since its equation is the one of a hyperboloid when viewed as a function of at least one loop momentum.

We now provide the multi-loop existence conditions for H-surfaces.
In the one-loop case, and in general when $|\mathcal{S}_i^{\mathbf{b},\sigma}|=2$, we have $\{j\}\equiv \{k \in \mathbf{b}\,|\,s^\mathbf{b}_{ik}\neq 0\}$ and the H-surface exists for real masses and loop momenta iff
\begin{equation}
(p_i^{0,\mathbf{b}})^2 - \vec{p}_i^{\,2}<(m_j-m_i)^2 \,,
\end{equation}
as already found in ref.~\cite{LTDRodrigoNumerical2017}.
In the case of $|\mathcal{S}_i^{\mathbf{b},\sigma}|>2$ and if exactly one H-surface sign differs from the others, whose index in $\mathbf{b}\cup\{i\}$ we label $\tilde{e}$, we define the following quantity:
\begin{equation}
\Delta M_i=\sum_{j\in \mathbf{b} }|s^\mathbf{b}_{ij}| (-1)^{\delta_{\tilde{e}j}} m_j + (-1)^{\delta_{\tilde{e}i}} m_i
\end{equation}
and the corresponding H-surface exists iff
\begin{equation}
\begin{cases}
\sigma_{\tilde{e}}p_i^{0,\mathbf{b}} < 0\ \mathrm{and}\ \Delta M_i<0 \ \mathrm{and}\ (p_i^{0,\mathbf{b}})^2 - \vec{p}_i^{\,2} < (\Delta M_i)^2 \\
\sigma_{\tilde{e}}p_i^{0,\mathbf{b}} > 0\ \mathrm{and}\ \Delta M_i>0 \ \mathrm{and}\ (p_i^{0,\mathbf{b}})^2 - \vec{p}_i^{\,2} > (\Delta M_i)^2\\
\sigma_{\tilde{e}}p_i^{0,\mathbf{b}} > 0\ \mathrm{and}\ \Delta M_i<0
\end{cases}
\end{equation}
or when the surface signs contain at least two positive and at least two negative members.

The singularities of dual integrands on H-surfaces cancel pairwise in their sum $\sum_{\mathbf{b} \in \mathcal{B}}
\mathrm{Res}_\mathbf{b}[f]$, due to a mechanism referred to as \emph{dual cancellations}~\cite{LTDRodrigoNumerical2017,Buchta:2014dfa,Aguilera-Verdugo:2019kbz}, independently of the regulator $\delta$.
We checked both numerically and analytically that eq.~\ref{eq:masterformula} maintains the dual cancellation pattern of H-surfaces also beyond two loops.

E-surfaces satisfy the following existence conditions for real masses and loop momenta:
\begin{equation}
(p_i^{0,\mathbf{b}})^2 - \vec{p}_i^{\,2} \geq \bigg( \sum_{j\in \mathbf{b} }|s^\mathbf{b}_{ij}|m_j + m_i \bigg)^2\ \mathrm{and}\ \ \sigma p_i^{0,\mathbf{b}}<0.
\end{equation}
We note that when the bound above is saturated, the E-surface is said to be \emph{pinched} and it corresponds to the location of physical soft and collinear singularities of the loop integral which would require dedicated local counterterms for its regularisation.

The singularities on existing E-surfaces must be regularised through a contour deformation satisfying its corresponding $\delta$-prescription.
We derive this prescription by writing the leading term of the Taylor expansion in $\delta$ of the imaginary part of $\Delta_i^{\sigma,\mathbf{b}}$:
\begin{equation}
\label{eq:imprescription}
\Im[\Delta_i^{\sigma,\mathbf{b}}]=-\frac{\delta}{2}\Bigg[\frac{\sigma}{E_i}+\sum_{j\in \mathbf{b}}\frac{s^{\mathbf{b}}_{ij} \sigma_j^{\mathbf{b}}}{E_j}\Bigg]+\mathcal{O}(\delta^2) \,.
\end{equation}
We note that for E-surfaces we have the definite sign $\sgn\Im[\Delta_i^{\sigma,\mathbf{b}}]=-\sigma$ independent of loop kinematics.
If no E-surface existence condition is satisfied, the integrand $\sum_{\mathbf{b} \in \mathcal{B}}
\mathrm{Res}_\mathbf{b}[f]$ has no singularities and it is therefore independent of the regulator $\delta$. In this case, the numerical integration can be performed without a contour deformation, a feature that has already been shown at one loop in ref.~\cite{LTDRodrigoNumerical2017} and two loops in ref.~\cite{Driencourt-Mangin:2019aix}.
A first preliminary result for a two-loop LTD integration using a contour deformation will be given in section~\ref{sec:outlook}.

\subsection{\label{sec:relation_previous_work} Discussion of previous work}

In ref.~\cite{LTDRodrigoMultiLoop2010} an alternative multi-loop LTD expression is derived by using distributional identities between dual and Feynman propagators instead of applying residue theorem. The main
distinction of the final expression lies in the $i\delta$-prescription of the dual propagator when diagrams beyond one loop are considered: in the case of ref.~\cite{LTDRodrigoMultiLoop2010}, the dual prescription is not equivalent to evaluating the on-shell conditions with complex energies $\sqrt{\vec{q}_i^{\ 2} + m^2 -\mathrm{i}\delta}$ and it is not a quantity independent of the order of integration in the loop variables, unless dual integrands with more on-shell conditions than loops are added.

The careful propagation of Feynman's causal prescription in the iterative approach discussed in the previous section is instrumental for obtaining a correct LTD expression for $n-$loop integrals. In the work of ref.~\cite{LTDWeinzierl2019}, an alternative LTD construction is presented, where an averaging procedure over all contour closures is considered, invoking the multidimensional residue theorem. The imaginary parts of each propagator are taken to be independent of each other throughout the induction proof, thereby not considering the interplay stemming from taking multiple on-shell conditions, such as the one reflected in eq.~\ref{eq:imaginaryinterplay}. \par

Our construction allows for arbitrarily choosing to close the contour of each energy variable either in the upper or lower complex half-plane. Using this, we have explicitly constructed the expression resulting from averaging over all possible contour choices and we find a different combination of residues than the one reported in ref.~\cite{LTDWeinzierl2019}. It thus appears that the aforementioned interplay in the determination of the sign of the imaginary part of each pole does not disappear upon this averaging procedure.
The integrand stemming from the direct combination of the integrands for each spanning tree given in ref.~\cite{LTDWeinzierl2019} does not reproduce our sum of dual integrands. Furthermore, we observe that this combined integrand does not realise dual cancellations.
We therefore conclude that the LTD expression presented in ref.~\cite{LTDWeinzierl2019} is incorrect beyond one loop, to the best of our understanding.

\section{\label{sec:numerics}Numerical application}
LTD has shown to yield promising results at one loop~\cite{LTDRodrigoNumerical2017} and has the advantage of not necessitating any computationally demanding symbolic treatment of the integrand and/or its numerator. This is different from sector decomposition techniques, which require building the Feynman representation of loop integrals together with the identification of sectors.
Moreover, integration in momentum space is particularly appealing for its optimal scaling with the number of contributing scales. Compared to the 4d momentum space integration method described in ref.~\cite{BeckerMultiLoop2012}, LTD has at least 5 advantages: (1) the dimension of the integration is reduced to 3 per loop, (2) a complex contour deformation only needs to be applied on bounded E-surfaces, (3) masses do not complicate the contour deformation much, (4) specific kinematical configurations can be integrated without any deformation, and (5) its singularity structure can directly be related to real-emission contributions~\cite{LTDRodrigoGauge2016}. 

In this work we are mostly interested in demonstrating LTD viability for numerical multi-loop computations and in assessing the validity of eq.~\ref{eq:masterformula}. Therefore, we apply LTD to loop integrals with external kinematics that do not yield singular E-surfaces, such that no complex contour deformation is required. This scenario offers a reliable numerical check of our LTD cut structures and of the numerical stability of the dual cancellations. Our implementation is a first important step towards handling loop integrals in the physical regime, which we briefly discuss in section~\ref{sec:outlook}.

\begin{table}[b]
\begin{adjustbox}{width=\columnwidth,center}
    \begin{tabular}{llllll}
    \toprule
        $G$ & Reference & Numerical LTD & N $[10^{6}]$ & [$\mu$s]\\ \hline
        
 & & & & \\[-1.2ex]
    
a)* &~\cite{Hirschi:2011pa} \hspace{0.04cm}$\,\mathrm{i}\,4.31638 \hspace{0.29cm}\cdot 10^{-7}$ & \hspace{0.03cm}$\,\mathrm{i}\,4.31637(19)\hspace{0.167cm}\cdot 10^{-7}$ & $110$ & $1.1$ \\[0.7ex]
    
   b)\phantom{*}
    &~\cite{Hirschi:2011pa}
    \hspace{0.06cm}$\,\mathrm{i}\,0.358640$ & \hspace{0.03cm}$\,\mathrm{i}\,0.358646(29)$ & 210 & $5.9$\\[0.7ex]
    
c)\phantom{*}
    &~\cite{Borowka:2017idc}
    \hspace{0.42cm}$1.1339(5)\hspace{0.02cm}\cdot 10^{-4}$ & \hspace{0.23cm}$1.133719(58) \cdot 10^{-4}$ & 5500 & $2.5$\\[0.7ex]
    
c)*
    &~\cite{Borowka:2017idc}
    \hspace{0.42cm}$4.398(1)\hspace{0.185cm}\cdot 10^{-8}$ & \hspace{0.25cm}$4.39825(17)\hspace{0.15cm}\cdot 10^{-8}$ & 5500 & $2.5$ \\[0.7ex]
    
d)*
    &~\cite{Borowka:2017idc}
    \hspace{0.42cm}$2.409(1)\hspace{0.18cm}\cdot 10^{-8}$ & \hspace{0.25cm}$2.40869(27)\hspace{0.15cm}\cdot 10^{-8}$ & 5500  & $3.5$ \\[0.7ex]

e)\phantom{*}
    &~\cite{Usyukina:1992jd}
    $-1.433521\hspace{0.13cm}\cdot 10^{-6}$& 
    $-1.4338(18)\hspace{0.3cm}\cdot 10^{-6}$ & 1500  & $27.4$ \\[0.7ex]

f)\phantom{*}
    &~\cite{Ruijl:2017cxj}
    $\hspace{0.12cm}\mathrm{i}\,5.26647 \hspace{0.27cm}\cdot 10^{-6}$& \hspace{0.06cm}$\,\mathrm{i}\,5.236(38)\hspace{0.47cm}\cdot 10^{-6}$ & 7000  & $3.3$ \\[0.7ex]
 
g)* &~\cite{Borowka:2017idc} \hspace{0.23cm}$\,\mathrm{i}\,1.7790(6)\hspace{0.02cm}\cdot 10^{-10}$ & \hspace{0.05cm}$\,\mathrm{i}\,1.77648(48)\hspace{0.16cm}\cdot 10^{-10}$ & 22000 & $11$\\[0.7ex]

h)\phantom{*} &~\cite{Ruijl:2017cxj}\hspace{0.02cm} $-8.36515\hspace{0.26cm}\cdot 10^{-8}$ & $-8.309(31)\hspace{0.47cm}\cdot 10^{-8}$ & 7000 & $15.8$\\
 
    \end{tabular}
\end{adjustbox}
    \caption{Comparison of our numerical LTD results for the topologies listed in table~\ref{tab:diagrams} against either the analytic result~\cite{Usyukina:1992jd,Ruijl:2017cxj} or an alternative numerical evaluation~\cite{Hirschi:2011pa,Borowka:2017idc}. A star indicates that internal lines are set massive. The columns labelled N and [$\mu$s] denote the Monte-Carlo statistics and timing per sample respectively. See details (incl. kinematic configurations) in ancillary material.}
    \label{tab:results}
    \end{table}

\begin{table}[h]
    \begin{tabular}{rcrcrcrc}
a) 
&
    \begin{tikzpicture}
    \begin{feynman}
    \tikzfeynmanset{every vertex={dot,minimum size=0.8mm}}
    \vertex (a1);
    
    \vertex[right=1cm of a1] (a3);
    
    \tikzfeynmanset{every vertex={empty dot,minimum size=0mm}}
    
    \vertex[right=0.5cm of a1] (a2);
    \vertex[above=0.5cm of a2] (a4);
    \vertex[below=0.5cm of a2] (a5);

    \vertex[left=0.433cm of a2] (b1);
    \vertex[left=0.25cm of a2] (b2);
    
    \vertex[right=0.433cm of a2] (b3);
    \vertex[right=0.25cm of a2] (b4);
    
    \vertex[left=0.433cm of a2] (d1);
    \vertex[left=0.25cm of a2] (d2);
    
    \vertex[right=0.433cm of a2] (d3);
    \vertex[right=0.25cm of a2] (d4);
    
    \tikzfeynmanset{every vertex={dot,minimum size=0.8mm}}
    
    \vertex[above=0.255cm of b1] (c1);
    \vertex[above=0.433cm of b2] (c2);
    
    \vertex[above=0.255cm of b3] (c3);
    \vertex[above=0.433cm of b4] (c4);
    
    \vertex[below=0.255cm of b1] (e1);
    \vertex[below=0.433cm of b2] (e2);
    
    \vertex[below=0.255cm of b3] (e3);
    \vertex[below=0.433cm of b4] (e4);
    
    \tikzfeynmanset{every vertex={empty dot,minimum size=0mm}}
    
    \vertex[left=0.15cm of c1] (q1);
    \vertex[left=0.15cm of c2] (q2);
    \vertex[left=0.15cm of e1] (q3);
    \vertex[left=0.15cm of e2] (q4);
    \vertex[left=0.15cm of a1] (q5);

    \vertex[right=0.15cm of c3] (p1);
    \vertex[right=0.15cm of c4] (p2);
    \vertex[right=0.15cm of e3] (p3);
    \vertex[right=0.15cm of e4] (p4);
    \vertex[right=0.15cm of a3] (p5);
    
        \diagram*[large]{	
        (a1)--[quarter left](a4) -- [quarter left](a3),
        (a5) --[quarter right](a3), 
        (a5)--[quarter left](a1),

        (c1) -- (q1),
        (c2) -- (q2),
        (e1) -- (q3), 
        (e2) -- (q4),
        (a1) -- (q5),

        (c3) -- (p1),
        (c4) -- (p2),
        (e3) -- (p3), 
        (e4) -- (p4),
        (a3) -- (p5),
        }; 
    \end{feynman}
    \end{tikzpicture}

&
b)
&

   \begin{tikzpicture}
    \begin{feynman}
    \tikzfeynmanset{every vertex={dot,minimum size=0.7mm}}

     \vertex (a1);

    \tikzfeynmanset{every vertex={empty dot,minimum size=0mm}}

    \vertex[right=1cm of a1] (a3);

    \vertex[right=0.5cm of a1] (a2);
    \vertex[above=0.5cm of a2] (a4);
    \vertex[below=0.5cm of a2] (a5);

    \vertex[right=0.4891cm of a2] (b1);
    \vertex[right=0.4568cm of a2] (b2);
    \vertex[right=0.4045cm of a2] (b3);
    \vertex[right=0.3346cm of a2] (b4);
    \vertex[right=0.25cm of a2] (b5);
    \vertex[right=0.1545cm of a2] (b6);
    \vertex[right=0.0523cm of a2] (b7);

    \vertex[right=0.6116cm of a2] (w1);
    \vertex[right=0.571cm of a2] (w2);
    \vertex[right=0.5056cm of a2] (w3);
    \vertex[right=0.4183cm of a2] (w4);
    \vertex[right=0.3125cm of a2] (w5);
    \vertex[right=0.1931cm of a2] (w6);
    \vertex[right=0.0654cm of a2] (w7);

    \vertex[above=0.13cm of w1] (x1);
    \vertex[above=0.2542cm of w2] (x2);
    \vertex[above=0.3674cm of w3] (x3);
    \vertex[above=0.4645cm of w4] (x4);
    \vertex[above=0.5413cm of w5] (x5);
    \vertex[above=0.5944cm of w6] (x6);
    \vertex[above=0.6216cm of w7] (x7);
    \vertex[below=0.13cm of w1] (xx1);
    \vertex[below=0.2542cm of w2] (xx2);
    \vertex[below=0.3674cm of w3] (xx3);
    \vertex[below=0.4645cm of w4] (xx4);
    \vertex[below=0.5413cm of w5] (xx5);
    \vertex[below=0.5944cm of w6] (xx6);
    \vertex[below=0.6216cm of w7] (xx7);

    \vertex[left=0.6116cm of a2] (v1);
    \vertex[left=0.571cm of a2] (v2);
    \vertex[left=0.5056cm of a2] (v3);
    \vertex[left=0.4183cm of a2] (v4);
    \vertex[left=0.3125cm of a2] (v5);
    \vertex[left=0.1931cm of a2] (v6);
    \vertex[left=0.0654cm of a2] (v7);

    \vertex[above=0.13cm of v1] (y1);
    \vertex[above=0.2542cm of v2] (y2);
    \vertex[above=0.3674cm of v3] (y3);
    \vertex[above=0.4645cm of v4] (y4);
    \vertex[above=0.5413cm of v5] (y5);
    \vertex[above=0.5944cm of v6] (y6);
    \vertex[above=0.6216cm of v7] (y7);
    \vertex[below=0.13cm of v1] (yy1);
    \vertex[below=0.2542cm of v2] (yy2);
    \vertex[below=0.3674cm of v3] (yy3);
    \vertex[below=0.4645cm of v4] (yy4);
    \vertex[below=0.5413cm of v5] (yy5);
    \vertex[below=0.5944cm of v6] (yy6);
    \vertex[below=0.6216cm of v7] (yy7);

    \vertex[left=0.4891cm of a2] (d1);
    \vertex[left=0.4568cm of a2] (d2);
    \vertex[left=0.4045cm of a2] (d3);
    \vertex[left=0.3346cm of a2] (d4);
    \vertex[left=0.25cm of a2] (d5);
    \vertex[left=0.1545cm of a2] (d6);
    \vertex[left=0.0523cm of a2] (d7);

    \vertex[right=0.625cm of a2] (qq1);
    \vertex[left=0.625cm of a2] (qq2);
    
    \tikzfeynmanset{every vertex={dot,minimum size=0.6mm}}

    \vertex[right=0.5cm of a2] (c0);

    \vertex[above=0.104cm of b1] (c1);
    \vertex[above=0.2034cm of b2] (c2);
    \vertex[above=0.2939cm of b3] (c3);
    \vertex[above=0.3716cm of b4] (c4);
    \vertex[above=0.433cm of b5] (c5);
    \vertex[above=0.4755cm of b6] (c6);
    \vertex[above=0.4973cm of b7] (c7);

    \vertex[above=0.104cm of d1] (e1);
    \vertex[above=0.2034cm of d2] (e2);
    \vertex[above=0.2939cm of d3] (e3);
    \vertex[above=0.3716cm of d4] (e4);
    \vertex[above=0.433cm of d5] (e5);
    \vertex[above=0.4755cm of d6] (e6);
    \vertex[above=0.4973cm of d7] (e7);

    \vertex[below=0.104cm of b1] (f1);
    \vertex[below=0.2034cm of b2] (f2);
    \vertex[below=0.2939cm of b3] (f3);
    \vertex[below=0.3716cm of b4] (f4);
    \vertex[below=0.433cm of b5] (f5);
    \vertex[below=0.4755cm of b6] (f6);
    \vertex[below=0.4973cm of b7] (f7);

    \vertex[below=0.104cm of d1] (g1);
    \vertex[below=0.2034cm of d2] (g2);
    \vertex[below=0.2939cm of d3] (g3);
    \vertex[below=0.3716cm of d4] (g4);
    \vertex[below=0.433cm of d5] (g5);
    \vertex[below=0.4755cm of d6] (g6);
    \vertex[below=0.4973cm of d7] (g7);
    
        \diagram*[large]{	
        (a1)--[quarter left](a4) -- [quarter left](a3),
        (a5) --[quarter right](a3), 
        (a5)--[quarter left](a1),
        
        (qq1) -- (c0),
        (qq2) -- (a1),
        
        (c1)-- (x1),
        (c2)-- (x2),
        (c3)-- (x3),
        (c4)-- (x4),
        (c5)-- (x5),
        (c6)-- (x6),
        (c7)-- (x7),
        
        (f1) -- (xx1),
        (f2) -- (xx2),
        (f3) -- (xx3),
        (f4) -- (xx4),
        (f5) -- (xx5),
        (f6) -- (xx6),
        (f7) -- (xx7),
        
        (e1)-- (y1),
        (e2)-- (y2),
        (e3)-- (y3),
        (e4)-- (y4),
        (e5)-- (y5),
        (e6)-- (y6),
        (e7)-- (y7),
        
        (g1) -- (yy1),
        (g2) -- (yy2),
        (g3) -- (yy3),
        (g4) -- (yy4),
        (g5) -- (yy5),
        (g6) -- (yy6),
        (g7) -- (yy7),
        
        }; 
    \end{feynman}
    \end{tikzpicture}

&
c)
&
 \begin{tikzpicture}
    \begin{feynman}

    \tikzfeynmanset{every vertex={empty dot,minimum size=0mm}}
    \vertex (a1);
    
    \vertex[right=1cm of a1] (a3);
    \vertex[right=0.5cm of a1] (a2);
    \vertex[above=0.5cm of a2] (a4);
    \vertex[below=0.5cm of a2] (a5);
    
    \vertex[left=0.433cm of a2] (b1);
    \vertex[right=0.433cm of a2] (b2);
    
    \vertex[above=0.25cm of b1] (c1);
    \vertex[above=0.25cm of b2] (c2);
    \vertex[below=0.25cm of b1] (d1);
    \vertex[below=0.25cm of b2] (d2);
    \vertex[above=0.3333cm of a2] (l1);
    \vertex[below=0.3333cm of a2] (l2);
    
    \vertex[left=0.15cm of c1] (ec1);
    \vertex[right=0.15cm of c2] (ec2);
    \vertex[left=0.15cm of d1] (ed1);
    \vertex[right=0.15cm of d2] (ed2);
    \vertex[right=0.15cm of l1] (el1);
    \vertex[right=0.15cm of l2] (el2);     
    
    \tikzfeynmanset{every vertex={dot,minimum size=0.8mm}}
    
    \vertex[above=0.25cm of b1] (c1);
    \vertex[above=0.25cm of b2] (c2);
    
    \vertex[below=0.25cm of b1] (d1);
    \vertex[below=0.25cm of b2] (d2);
    
    \vertex[above=0.3333cm of a2] (l1);
    \vertex[below=0.3333cm of a2] (l2);
    
        \diagram*[large]{	
        (a1)--[quarter left](a4) -- [quarter left](a3),
        (a5) --[quarter right](a3), 
        (a5)--[quarter left](a1),
        (a4) -- (a5),
        
        (c1) -- (ec1),
        (c2) -- (ec2),
        (d1) -- (ed1),
        (d2) -- (ed2),
        (l1) -- (el1),
        (l2) -- (el2),

        }; 
    \end{feynman}
    \end{tikzpicture}

&
d)
&

    \begin{tikzpicture}
    \begin{feynman}

    \tikzfeynmanset{every vertex={empty dot,minimum size=0mm}}
    \vertex (a1);
    
    \vertex[right=1cm of a1] (a3);
    \vertex[right=0.5cm of a1] (a2);
    \vertex[above=0.5cm of a2] (a4);
    \vertex[below=0.5cm of a2] (a5);
    
    \vertex[left=0.433cm of a2] (b1);
    \vertex[right=0.433cm of a2] (b2);
    
    \vertex[above=0.25cm of b1] (c1);
    \vertex[above=0.25cm of b2] (c2);
    \vertex[below=0.25cm of b1] (d1);
    \vertex[below=0.25cm of b2] (d2);
    \vertex[below=0.2cm of a4] (l1);
    \vertex[below=0.4cm of a4] (l2);
    \vertex[below=0.6cm of a4] (l3);
    \vertex[below=0.8cm of a4] (l4);
    
    \vertex[left=0.15cm of c1] (ec1);
    \vertex[right=0.15cm of c2] (ec2);
    \vertex[left=0.15cm of d1] (ed1);
    \vertex[right=0.15cm of d2] (ed2);
    \vertex[right=0.15cm of l1] (el1);
    \vertex[right=0.15cm of l2] (el2);
    \vertex[right=0.15cm of l3] (el3);
    \vertex[right=0.15cm of l4] (el4);

    \tikzfeynmanset{every vertex={dot,minimum size=0.8mm}}
    
    \vertex[above=0.25cm of b1] (c1);
    \vertex[above=0.25cm of b2] (c2);
    
    \vertex[below=0.25cm of b1] (d1);
    \vertex[below=0.25cm of b2] (d2);
    
    \vertex[below=0.2cm of a4] (l1);
    \vertex[below=0.4cm of a4] (l2);
    \vertex[below=0.6cm of a4] (l3);
    \vertex[below=0.8cm of a4] (l4);
    
        \diagram*[large]{	
        (a1)--[quarter left](a4) -- [quarter left](a3),
        (a5) --[quarter right](a3), 
        (a5)--[quarter left](a1),
        (a4) -- (a5),
    
        (c1) -- (ec1),
        (c2) -- (ec2),
        (d1) -- (ed1),
        (d2) -- (ed2),
        (l1) -- (el1),
        (l1) -- (el1),
        (l2) -- (el2),
        (l3) -- (el3),
        (l4) -- (el4),
        }; 
    \end{feynman}
    \end{tikzpicture}
   \\
   \\
e)

&
        \begin{tikzpicture}
            \begin{feynman}

            \tikzfeynmanset{every vertex={dot,minimum size=1mm}}

            \tikzfeynmanset{every vertex={empty dot,minimum size=0mm}}
             \vertex (a1);
            \vertex[right=0.25cm of a1] (a2);
            \vertex[above=0.5cm of a2] (a4);
            \vertex[below=0.5cm of a2] (a5);

            \vertex[right=1cm of a1] (d1);

            \vertex[right=0.25cm of a4] (b1);
            \vertex[right=0.25cm of a5] (b2);
            \vertex[right=0.25cm of b1] (c1);
            \vertex[right=0.25cm of b2] (c2);
            
            \vertex[above=0.5cm of a1] (a7);
            \vertex[below=0.5cm of a1] (a8);
            \vertex[above=0.5cm of d1] (a9);
            \vertex[below=0.5cm of d1] (a10);
            
            \vertex[left=0.15cm of a7] (a11);
            \vertex[left=0.15cm of a8] (a12);
            \vertex[right=0.15cm of a9] (a13);
            \vertex[right=0.15cm of a10] (a14);
            
            \tikzfeynmanset{every vertex={dot,minimum size=0.8mm}}

            \vertex[above=0.5cm of a1] (e1);
            \vertex[below=0.5cm of a1] (e2);    
            
            \vertex[above=0.5cm of d1] (f1);
            \vertex[below=0.5cm of d1] (f2);

                \diagram*[large]{	
                (e1)--(a4), 
                (a5)--(e2),
		(e1)--(e2),

                (a4) -- (a5),
                (a4) -- (b1),
                (a5) -- (b2),
                (b1)--(b2),
                (b1)--(c1),
                (b2)--(c2),
                (c1)--(c2),

                (c1) --  (f1),
                (f2) -- (c2),
		        (f1)--(f2),
		        
		        (a7) -- (a11),
		        (a8) -- (a12),
		        (a9) -- (a13),
		        (a10) -- (a14),
                }; 
            \end{feynman}
    \end{tikzpicture}

&
f)
&

    \begin{tikzpicture}
    \begin{feynman}
    
    \tikzfeynmanset{every vertex={empty dot,minimum size=0mm}}
    
    \vertex (a1);
    
    \vertex[below=0.5cm of a1] (a3);
    \vertex[below=0.5cm of a3] (b1);
    \vertex[below=0.35355cm of a3] (a5);
    \vertex[left=0.35355cm of a5] (a6);
    
    \vertex[left=0.5cm of a3] (a7);
    \vertex[left=0.15cm of a7] (a8);
    
    \vertex[right=0.5cm of a3] (a9);
    \vertex[right=0.15cm of a9] (a10);

    \tikzfeynmanset{every vertex={dot,minimum size=0.8mm}}
    \vertex[left=0.5cm of a3] (a2);
    \vertex[right=0.5cm of a3] (a4);

        \diagram*[large]{	
        (a1)--[quarter left](a4) -- [quarter left](b1) -- [quarter left](a2)-- [quarter left](a1),
        (a3) -- (a1),
        (a3) -- (a6),
        (a3) -- (a4),
        (a7) -- (a8),
        (a9) -- (a10),
        }; 
    \end{feynman}
    \end{tikzpicture}

&
g)
&

    \begin{tikzpicture}
    \begin{feynman}

    \tikzfeynmanset{every vertex={empty dot,minimum size=0mm}}
    
    \vertex (a1);
    
    \vertex[below=0.5cm of a1] (a3);
    \vertex[below=0.5cm of a3] (b1);
    \vertex[below=0.35355cm of a3] (a5);
    \vertex[left=0.35355cm of a5] (a6);
    
    \vertex[below=0.17678cm of a3] (b2);
    
    \vertex[below=0.14645cm of a1] (b3);
    
    \vertex[below=0.35355cm of a3] (b4);
    
    \vertex[left=0.5cm of a3] (a7);
    \vertex[left=0.15cm of a7] (a8);
    
    \vertex[right=0.5cm of a3] (a9);
    \vertex[right=0.15cm of a9] (a10);
    
    \vertex[left=0.35355cm of b3] (a11);
    \vertex[left=0.15cm of a11] (a12);
    
    \vertex[below=0.25cm of a1] (a13);
    \vertex[right=0.15cm of a13] (a14);
    
    \vertex[right=0.35355cm of b4] (a15);
    \vertex[right=0.15cm of a15] (a16);
    
    \vertex[left=0.17678cm of b2] (a17);
    \vertex[right=0.15cm of a17] (a18);
    
    \tikzfeynmanset{every vertex={dot,minimum size=0.8mm}}
    \vertex[left=0.5cm of a3] (a2);
    \vertex[right=0.5cm of a3] (a4);
    
    \vertex[left=0.17678cm of b2] (c1);
    \vertex[below=0.25cm of a1] (c2);
    
    \vertex[left=0.35355cm of b3] (c3);
    \vertex[right=0.35355cm of b4] (c4);
    
        \diagram*[large]{	
        (a1)--[quarter left](a4) -- [quarter left](b1) -- [quarter left](a2)-- [quarter left](a1),
        (a3) -- (a1),
        (a3) -- (a6),
        (a3) -- (a4),
        (a7) -- (a8),
        (a9) -- (a10),
        (a11) -- (a12),
        (a13) -- (a14),
        (a15) -- (a16),
        (a17) -- (a18),
        }; 
    \end{feynman}
    \end{tikzpicture}

&
h)
&

    \begin{tikzpicture}
    \begin{feynman}

    \tikzfeynmanset{every vertex={empty dot,minimum size=0mm}}
    
    \vertex (a1);
    
    \vertex[below=0.5cm of a1] (a3);
    
    \vertex[right=0.08cm of a3] (cross1);
    \vertex[left=0.08cm of a3] (cross2);
    
    \vertex[below=0.5cm of a3] (b1);
    
    \vertex[below=0.15cm of a1] (c1);
    \vertex[right=0.35cm of a3] (c2);
    
    \vertex[below=0.35355cm of a3] (a5);
    
    \vertex[left=0.5cm of a3] (a2);
    
    \vertex[right=0.5cm of a3] (a9);
    \vertex[right=0.15cm of a9] (a10);
    
    \vertex[left=0.35355cm of a5] (a11);
    \vertex[left=0.15cm of a11] (a12);
    
    \tikzfeynmanset{every vertex={dot,minimum size=0.8mm}}
    
    \vertex[right=0.5cm of a3] (a4);
    \vertex[left=0.35355cm of a5] (a6);

        \diagram*[large]{	
        (a1)--[quarter left](a4) -- [quarter left](b1) -- [quarter left](a2)-- [quarter left](a1),
        (a1) -- (c1),
        (c1) -- (c2),
        (c2) -- (a4),
        (c2) -- (cross1),
        (a2) -- (cross2),
        (cross1) --[white] (cross2),	
        (c1) -- (b1),
        (a9) -- (a10),
        (a11) -- (a12),
        }; 
    \end{feynman}
    \end{tikzpicture}

    \end{tabular}
    \caption{Scalar loop diagrams considered in our numerical validation. A small line attached to a dotted vertex denotes an insertion of an external momentum. Graph b) has 30 legs.}
    \label{tab:diagrams}
\end{table}

\begin{figure}[H]
	\centering
	\includegraphics[trim=98 245 45 200, clip,scale=0.50]{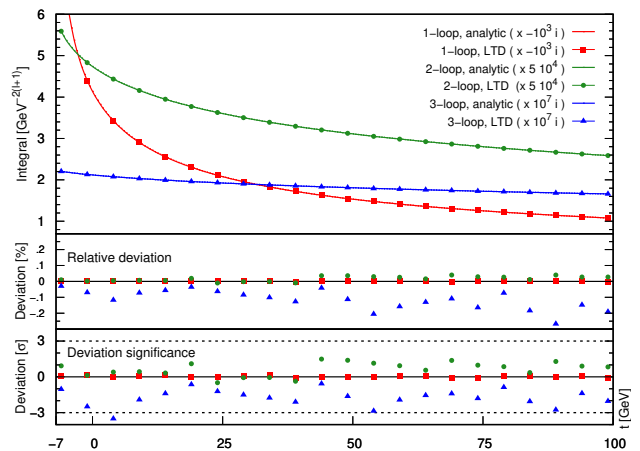}
	\caption{\label{fig:BoxesScan}
Numerical LTD results obtained for the scalar massless one-, two- and three-loop ladder box diagram with external kinematics satisfying (in GeV$^2$) $p_1^2=-5$, $p_{i=2,3,4}^2=s=-1$ and values of the Mandelstam invariant $t$ ranging from $t=-7$ (loop threshold) to $t=100$. The analytic results are taken from ref.~\cite{Usyukina:1992jd}.
}
\end{figure}

We selected eight very different loop topologies, displayed in table~\ref{tab:diagrams}, to showcase the generality of the method.
We report our results in table~\ref{tab:results} and figure~\ref{fig:BoxesScan}, with additional information (such as the exact input kinematics) given as ancillary material to ensure reproducibility of our work. The reference results are taken from the analytic expression for the four-point integrals~\cite{Usyukina:1992jd}, from {\sc\small Forcer}~\cite{Ruijl:2017cxj} for two-point integrals, from {\sc\small MadLoop}~\cite{Hirschi:2011pa, Alwall:2014hca} for the decagon and triacontagon and {\sc\small PySecDec}~\cite{Borowka:2017idc} for the six- and eight-point integrals (in which case the numerical error is also reported).
We find perfect agreement in all cases, but note that scalar integrals whose superficial degree of UV divergence is -2 (\ref{tab:diagrams}.f and \ref{tab:diagrams}.h) are numerically more challenging. This is made manifest for example when comparing LTD results obtained for the loops \ref{tab:diagrams}.f and \ref{tab:diagrams}.g. We find no notable sensitivity of the numerical convergence to the external momenta multiplicity, internal masses or non-planarity of the loop graph.

For all eight benchmark loop integrals, we have explicitly verified that dual cancellations hold by sampling points on the H-surfaces for which we found that the sum of dual integrands is regular. It is important to monitor numerical stability when probing points close to such surfaces, as dual cancellations occur by cancelling large summands. We monitor this stability by testing the invariance of dual integrands under rotation of the spatial parts of the loop momenta integrated over. The more challenging loop integrals required a custom numerical stability rescue system that promotes the floating point arithmetic accuracy to quadruple precision when needed (which is about a factor 30 slower). We note however that the introduction of a complex contour deformation mitigates the numerical severity of dual cancellations.

Since we are mostly interested in verifying our method at this stage, we stress that no effort was made to fine-tune the integrator, sample statistics or loop momenta parametrisations. Sizeable improvements can be expected from considering techniques similar to the ones described in ref.~\cite{Becker:2012aqa}. Similarly to what was found in ref.~\cite{LTDRodrigoNumerical2017}, we observe that the {\sc\small Cuhre} integrator offers significantly better convergence at one loop. However, we find that it performs much worse than {\sc\small Vegas} at higher loops. For uniformity, we restricted ourselves to using the {\sc\small Vegas} integrator for producing the results of table~\ref{tab:results}. Our implementation is written in the {\sc\small Rust} language, with {\sc\small Python} bindings, and interfaces to the {\sc\small Cuba}~\cite{Cuba2005} library and {\sc\small Vegas3.4}~\cite{Lepage:1977sw} for performing the adaptive Monte-Carlo integration.

\section{\label{sec:outlook} General kinematics}

The LTD expression of eq.~\ref{eq:masterformula}, evaluated at external kinematics relevant for computing physical scattering amplitudes, typically features singular E-surfaces. The corresponding singularities then require a complex contour deformation of the spatial part of the loop variables, constructed so as to satisfy the LTD prescription associated to the surface, presented expanded at the first order in $\delta$ in \eqref{eq:imprescription}.
In sect.~\ref{sec:surfaces}, we found that the sign of the imaginary part of the defining equation of E-surfaces reads:
\begin{equation}\label{prescr_ellipsoids}
\sgn \Im[\Delta_i^{\mathbf{b},\sigma}]=-\sigma\,.
\end{equation}

We now aim at constructing a contour deformation that satisfies the causality constraints implied by the $\mathrm{i} \delta$ prescription.
Given its parametrisation $\vec{k}_l^{\mathbb{C}}=\vec{k}_l+\mathrm{i}\vec{K}_l, \ l\in\{1,...,n\}$, one has that $\vec{q}_j\rightarrow \vec{q}_j+\mathrm{i}\vec{\kappa}_j$, $\forall j\in\mathbf{b}$, and the imaginary part of every other propagator momentum can be expressed as a linear combination of $\{\vec{\kappa}_j\}_{j\in\mathbf{b}}$. This results in $\Delta_{i}^{\mathbf{b},\sigma}$ acquiring an imaginary part, 
\begin{equation}
\label{eq:imdeformation}
\Im[\Delta_i^{\mathbf{b},\sigma}]=\sum_{j\in \mathbf{b}}s_{ij}^{\mathbf{b}}\vec{\kappa}_j\cdot\Bigg(\frac{ \sigma_j\vec{q}_j}{2E_j}+\frac{\sigma\vec{q}_i}{2E_i}\Bigg),
\end{equation}
in the first order truncation of the expansion in $\sqrt{\sum_j |s^\mathbf{b}_{ij}| \vec{\kappa}_j^2} $. For E-surfaces, this simplifies to
\begin{equation}
\label{eq:deformation}
\Im[\Delta_i^{\mathbf{b},\sigma}]=\frac{\sigma}{2}\sum_{j\in \mathbf{b}}|s^\mathbf{b}_{ij}|\vec{\kappa}_j\cdot \vec{v}_{i,j}^{\,\mathbf{b}}, \hspace{0.5cm} \vec{v}_{i,j}^{\,\mathbf{b}}=\frac{\vec{q}_j}{E_j}+\frac{s_{ij}^{\mathbf{b}}\vec{q}_i}{E_i},
\end{equation}

which can be matched with \eqref{prescr_ellipsoids} on individual E-surfaces by just setting $\vec{\kappa}_j\propto -\vec{v}_{i,j}^{\,\mathbf{b}}, \ \forall j \in\mathbf{b}$ such that $s_{ij}^{\mathbf{b}}\neq 0$. Now let $\text{H}(w)=\{v\in\mathbb{R}^3 \ | \ w\cdot v>0\}$. One can observe that for every value of the loop variables
\begin{equation}\label{intersection}
\vec{q}_j\in\bigcap_{l\in\mathcal{E}_j^{\mathbf{b}}}  \text{H}(\vec{v}_{l,j}^{\,\mathbf{b}}), \ \hspace{0.15cm} \forall j\in\mathbf{b},
\end{equation}
where $\mathcal{E}_j^{\mathbf{b}}=\{l\in\mathbf{e}\setminus \mathbf{b}\ \vert\  |s_{lj}^{\mathbf{b}}|\vec{v}^{\mathbf{\,b}}_{l,j}\neq \vec{0}  \}$. Thus, $\vec{\kappa}_j\propto-\vec{q}_j$, $\forall j\in\mathbf{b}$ satisfies the prescription on arbitrarily many E-surfaces associated to the same loop momentum basis $\mathbf{b}$, including on their intersection. Indeed, $\vec{q}_j$ has positive projection on all non zero $|s_{ij}|\vec{v}_{i,j}^{\mathbf{\,b}}$ which might appear as summands in \eqref{eq:deformation}. This fails only if there exists an E-surface $S_{l}^{\mathbf{b},\sigma}$ such that $|s_{lj}|\vec{v}_{l,j}^{\,\mathbf{b}}=\vec{0}, \ \forall j\in\mathbf{b}$, which would correspond to a pinched surface necessitating a soft/collinear regulator or subtraction. Finally, the intersections of two surfaces $S_{a}^{\mathbf{b},\sigma}$ and $S_{b}^{\mathbf{\tilde{b}},\tilde{\sigma}}$ with $\mathbf{b}\neq\mathbf{\tilde b}$ would lie on dual cancelling surfaces, and are thus not singular in the sum of dual integrands.
We stress that the above does not provide a complete recipe for building an overall continuous deformation direction $\vec{K}_l$ satisfying all causal constraints and common to all dual integrands so as to preserve dual cancellations. This requires a (numerically efficient) strategy for interpolating between the deformation directions identified in eq.~\eqref{intersection} for each group of E-surfaces $\mathcal{E}_j^{\mathbf{b}}$. Additionally, special care must be taken when setting the normalisation of the resulting deformation vector $\vec{K}_l$.

We conclude this section by presenting a first two-loop numerical result from applying LTD to a double-box topology that requires a deformation around its 13 distinct E-surfaces. We set the external kinematics identical to those of the benchmark point chosen in ref.~\cite{BeckerMultiLoop2012} and also report them in the ancillary material. Using {\sc\small Vegas3.4} with 105M Monte-Carlo samples, we obtained $-5.877(55)\cdot 10^{-14}$, which stands within 1\% of the analytical result $-5.8973\cdot 10^{-14}$.
We will provide a general and numerically efficient contour deformation for multi-loop LTD in an upcoming publication.

\section{\label{sec:conclusion}Conclusion}
We derived a novel expression for multi-loop LTD, that involves taking as many on-shell conditions as there are loops. We demonstrated its potential for numerical integration by applying it to eight finite scalar multi-loop topologies. Additionally, we gave a first result of a contour deformation at two loops, showing that LTD can be used for computing integrals with physical kinematics as well.

Multi-loop LTD is a promising approach from both an analytic and a numerical perspective. One challenging possibility is to directly combine virtual and real-emission unresolved degrees of freedom, as already explored at one loop in ref.~\cite{LTDRodrigoGauge2016}, allowing for their joint integration with fewer or no counterterms. For numerical integration, LTD gives the advantage of reducing the number of dimensions to three per loop momentum, and confines singular surfaces one must deform around to a bounded region.

Our future work concerns extending the application of LTD to diagrams and loop amplitudes featuring (1) complicated overlaps of E-surfaces requiring a general contour deformation, and (2) UV and IR divergences, by designing local subtraction counterterms that leverage known factorisation properties, such as the ones introduced in ref.~\cite{Anastasiou:2018rib}.

\section*{Acknowledgements}
We would like to thank Andrea Pelloni and Babis Anastasiou for useful discussions.
This project has received funding from the European Research Council (ERC) under grant agreement No 694712 (PertQCD) and SNSF grant No 179016. Numerical results presented in this letter used computational resources from the Piz Daint cluster, administered by the Swiss National Supercomputing Centre (CSCS).

\bibliographystyle{apsrev4-1}
\bibliography{biblio}

\appendix
\onecolumngrid
\date{\nodate}

\section{Explicit example of a three-loop cut structure}

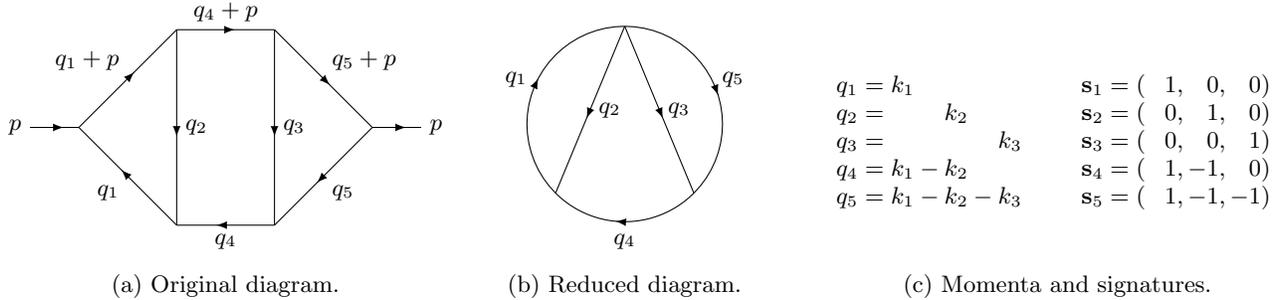
\begin{figure}[h!]
\centering
\begin{subfigure}[b]{0.35\textwidth}
\centering
\begin{tikzpicture}[scale=1.3]
\coordinate (A) at (0,0);
\coordinate (B) at (0.5,0);
\coordinate (C) at (1.5,1);
\coordinate (D) at (2.5,1);
\coordinate (E) at (3.5,0);
\coordinate (F) at (4,0);

\coordinate (G) at (1.5,-1);
\coordinate (H) at (2.5,-1);

\draw[->-] node[left]{$p$} (A) -- (B);
\draw[->-] (B) -- node[above left]{$q_1+p$} (C);
\draw[->-] (C) -- node[above]{$q_4+p$} (D);
\draw[->-] (D) -- node[above right]{$q_5+p$} (E);
\draw[->-] (E) -- (F) node[right]{$p$};
\draw[->-] (G) -- node[below left]{$q_1$} (B);
\draw[->-] (H) -- node[below]{$q_4$} (G);
\draw[->-] (E) -- node[below right]{$q_5$} (H);
\draw[->-] (C) -- node[right]{$q_2$} (G);
\draw[->-] (D) -- node[right]{$q_3$} (H);
\end{tikzpicture}
\caption{Original diagram.}
\label{fig:example_a}
\end{subfigure}
\begin{subfigure}[b]{0.23\textwidth}
\centering
\begin{tikzpicture}[scale=1.3]

\draw[domain=225:90,variable=\x,->-] plot ({cos(\x)},{sin(\x});
\node at (157.5:1.2){$q_1$};

\draw[domain=45:135,variable=\x,->-]  plot ({cos(\x)},{-sin(\x});
\node at (270:1.2){$q_4$};

\draw[domain=90:-45,variable=\x,->-]  plot ({cos(\x)},{sin(\x});
\node at (22.5:1.2){$q_5$};

\draw[->-] (90:1) -- node[right]{$q_2$} (225:1);
\draw[->-] (90:1) -- node[right]{$q_3$} (-45:1);

\end{tikzpicture}
\caption{Reduced diagram.}
\label{fig:example_b}
\end{subfigure}
\begin{subfigure}[b]{0.4\textwidth}
\centering
\begin{tabular}{ l r }
$q_1 = k_1$ & \qquad
$\mathbf{s}_1 = (\phantom{+}1,\phantom{+}0,\phantom{+}0)$\\
$q_2 = \phantom{k_1{}+{}}k_2$ & \qquad
$\mathbf{s}_2 = (\phantom{+}0,\phantom{+}1,\phantom{+}0)$\\
$q_3 = \phantom{k_1{}+{}k_2{}+{}}k_3$ & \qquad
$\mathbf{s}_3 = (\phantom{+}0,\phantom{+}0,\phantom{+}1)$\\
$q_4 = k_1 - k_2$ & \qquad
$\mathbf{s}_4 = (\phantom{+}1,-1,\phantom{+}0)$\\
$q_5 = k_1 - k_2 - k_3$ & \qquad
$\mathbf{s}_5 = (\phantom{+}1,-1,-1)$\\
\end{tabular}
\vspace*{6mm}
\caption{Momenta and signatures.}
\label{fig:example_c}
\end{subfigure}

\caption{A 2-point 3-loop ladder diagram and its reduced diagram, obtained by merging propagator lines that share identical signatures and removing external legs.}
\label{fig:example}
\end{figure}

This section provides an example of the application of our LTD expression for a three-loop topology and it illustrates how the \emph{cut structure} output by the \textsc{\small Python} script given in ancillary material is intended to be used. \par

Two loop momentum bases share the same cut structures if their set of signatures is identical, in other words if their defining loop momenta only differ by constant shifts.
This implies that propagators sharing the same signature can be combined into a single list that we refer to here as a \emph{loop line}, as done in ref~\cite{LTDRodrigoOrigin2008}.
A graphical equivalent of this procedure corresponds to constructing a \emph{reduced} version of a loop diagram, where propagators with identical signatures are merged into a single loop line (see fig.~\ref{fig:example_b}). \par

For a given reduced diagram (encoded as a list of loop momentum signatures), we can write the cut structure corresponding to each specific momentum basis (or equivalently: spanning tree) $\mathbf{b}_i$ as a list $\boldsymbol\Sigma_i$ containing as many elements as there are loop lines.
At $n$ loops, $n$ of these elements are the members of ${\boldsymbol\sigma}^{\mathbf{b}_i}$ and the remaining ones are set to $0$ and serve to identify this particular spanning tree of the reduced diagram. \par

The cut structures of the diagram in fig.~\ref{fig:example_a} (with $30$ spanning trees) can therefore be obtained from the cut structures of the (simpler) reduced diagram in fig.~\ref{fig:example_b} (with $8$ spanning trees). As we see in fig.~\ref{fig:example_c}, both graphs have the same set of signature vectors. \par

For the reduced diagram in fig.~\ref{fig:example_b} and when opting to close the contour of each energy integration in the lower-half complex plane, the \textsc{\small Python} script provided will return the following cut structures $\boldsymbol\Sigma_i$, each specifying all at once the spanning tree to consider and the cut energy signs corresponding to it:
\begin{align}
\boldsymbol\Sigma_1 &= (\phantom{+}1, \phantom{+}1, \phantom{+}1, \phantom{+}0, \phantom{+}0), \nonumber\\
\boldsymbol\Sigma_2 &= (\phantom{+}1, -1, \phantom{+}0, \phantom{+}0, -1), \nonumber\\
\boldsymbol\Sigma_3 &= (\phantom{+}1, \phantom{+}0, \phantom{+}1, -1, \phantom{+}0), \nonumber\\
\boldsymbol\Sigma_4 &= (\phantom{+}1, \phantom{+}0, \phantom{+}1, \phantom{+}0, -1), \nonumber\\
\boldsymbol\Sigma_5 &= (\phantom{+}1, \phantom{+}0, \phantom{+}0, \phantom{+}1, -1), \nonumber\\
\boldsymbol\Sigma_6 &= (\phantom{+}0, \phantom{+}1, \phantom{+}1, \phantom{+}1, \phantom{+}0), \nonumber\\
\boldsymbol\Sigma_7 &= (\phantom{+}0, \phantom{+}1, \phantom{+}1, \phantom{+}0, \phantom{+}1), \nonumber\\
\boldsymbol\Sigma_8 &= (\phantom{+}0, \phantom{+}1, \phantom{+}0, \phantom{+}1, -1).
\end{align}
There are as many cut structures as there are spanning trees. For each cut structure, the element $\sigma \in \{\pm1,0\}$ at position $i$ denotes that the loop line corresponding to $\mathbf{s}_i$ is either cut with a positive/negative ($\sigma=+1/-1$) energy solution or not cut at all ($\sigma=0$). \par

When a loop line propagator is cut, one must consider one residue per propagator building this loop line. The energy solutions of these residues are then all equal up to constant shifts dictated by the energy component of the external momenta inserted along this loop line.
When multiple loop lines are cut, all possible combinations of propagator cuts of each loop line must be considered. For example in the reduced diagram of fig.~\ref{fig:example_b}, cutting the loop lines $q_1$, $q_4$ and $q_5$ (corresponding to the cut structure $\boldsymbol\Sigma_5$) would yield $2\times2\times2=8$ residues, while $\boldsymbol\Sigma_1$ would yield only $2\times1\times1=2$ residues.
Alternatively, one can also opt to generate the cut structures of the original diagram of fig.~\ref{fig:example_a} by treating each propagator independently. This would yield $30$ cut structures, each in the form of a list containing $8$ elements (one per loop line), each this time corresponding to a single residue.\par

We now make explicit the notation $\{q_j^0=\sigma^{\mathbf{b}}_j E_j\}_{j\in \mathbf{b}}$ by writing, in the case of massless propagators, what are two of the four energy solutions defining the residues corresponding to the cut structure $\boldsymbol\Sigma_2$. These are the loop momentum energy configurations solving eq.~\eqref{system-residues} of the main text.
A first residue stems from the energy solutions arising from cutting the propagators with momenta $q_1$, $q_2$ and $q_5$
\begin{align}
k_1^0 &= \phantom{+}\sqrt{\vec{k}_1^2- \mathrm{i}\delta} \nonumber\\
k_2^0 &= -\sqrt{\vec{k}_2^2- \mathrm{i}\delta} \nonumber\\
k_1^0 - k_2^0 - k_3^0 &= -\sqrt{(\vec{k}_1 - \vec{k}_2 -\vec{k}_3)^2- \mathrm{i}\delta},
\end{align}
while a second one, for the same cut structure  $\boldsymbol\Sigma_2$, comes from cutting the propagators with momenta $q_1+p$, $q_2$ and $q_5$
\begin{align}
k_1^0 + p^0 &= \phantom{+}\sqrt{(\vec{k}_1+\vec{p})^2- \mathrm{i}\delta} \nonumber\\
k_2^0 &= -\sqrt{\vec{k}_2^2- \mathrm{i}\delta} \nonumber\\
k_1^0 - k_2^0 - k_3^0 &= -\sqrt{(\vec{k}_1 - \vec{k}_2 -\vec{k}_3)^2- \mathrm{i}\delta}
\end{align}
and similarly for the remaining two residues.

\newpage
\section{External kinematics considered for our numerical results}

\subsection{Topology a)* and b): 1-loop deca- and triacontagon}

The following diagram describes the momentum flow for the two one-loop scalar integrals considered in our numerical validation section:
\begin{center}
\begin{tikzpicture}
\pgfmathsetmacro{\r}{1.5}
\pgfmathsetmacro{\n}{6}
\pgfmathsetmacro{\angle}{360/\n}
\foreach \y in {1,...,\n}{
	\pgfmathsetmacro{\start}{\y*\angle}
	\pgfmathsetmacro{\stop}{(\y+1)*\angle}
	\pgfmathsetmacro{\mid}{(\y+0.5)*\angle}
	\ifthenelse{\y=3}{
		\draw [dotted] (\start:\r*1.5) -- (\start:\r);
		\draw [domain=\start:\stop,dotted] plot ({\r*cos(\x)}, {\r*sin(\x)});}{
		\ifthenelse{\y=4}{
			\draw [dotted] (\start:\r*1.5) -- (\start:\r);}{
			\draw [->-] (\start:\r*1.5) -- (\start:\r);}
		\draw [domain=\start:\stop,->-] plot ({\r*cos(\x)}, {\r*sin(\x)});}
}
\node at (1*\angle:\r*1.7) { $p_{1}$};
\node at ({(1+0.5)*\angle}:\r*0.7) { $k_{1}$};
\node at (2*\angle:\r*1.7) { $p_{2}$};
\node at ({(2+0.5)*\angle}:\r*0.7) { $k_{2}$};

\node at ({(4+0.5)*\angle}:\r*0.7) { $k_{n-2}$};
\node at (5*\angle:\r*1.7) { $p_{n-1}$};
\node at ({(5+0.5)*\angle}:\r*0.7) { $k_{n-1}$};
\node at (6*\angle:\r*1.7) { $p_{n}$};
\node at ({(6+0.5)*\angle}:\r*0.7) { $k_n$};

\end{tikzpicture}
\end{center}

Kinematics for the decagon a)*:

\begin{footnotesize}
  \begin{align*}
    [GeV]         &\phantom{=\textrm{( 0}}E&&           \phantom{\textrm{,0}}p_x                            &&\phantom{\textrm{,0}}p_y                           &&\phantom{\textrm{,0}}p_z                           &&\\
    p_1            &=\textrm{( \phantom{-}4.999749993750}\cdot 10^{\textrm{-1}}&&                           \textrm{, \phantom{-}0}                               &&\textrm{, \phantom{-}0}                               &&\textrm{, \phantom{-}2.5}                           &&\textrm{)}\\    
    p_2            &=\textrm{( -4.999749993750}\cdot 10^{\textrm{-1}}&&                           \textrm{, \phantom{-}0}                               &&\textrm{, \phantom{-}0}                               &&\textrm{, \phantom{-}2.5}                                            &&\textrm{)}\\
    p_3            &=\textrm{( \phantom{-}5.319269421240}\cdot 10^{\textrm{-2}}&&\textrm{, \phantom{-}7.630500824806}\cdot 10^{\textrm{-3}}&&\textrm{, -4.191452656443}\cdot 10^{\textrm{-2}}&&\textrm{, -3.416692405504}\cdot 10^{\textrm{-1}}                  &&\textrm{)}\\
    p_4            &=\textrm{( -7.418366402852}\cdot 10^{\textrm{-3}}&&\textrm{, \phantom{-}4.793330102618}\cdot 10^{\textrm{-2}}&&\textrm{, \phantom{-}1.092200259939}\cdot 10^{\textrm{-1}}&&\textrm{, -5.980519957635}\cdot 10^{\textrm{-1}}                  &&\textrm{)}\\
    p_5            &=\textrm{( -1.031283270712}\cdot 10^{\textrm{-2}}&&\textrm{, \phantom{-}5.274052562103}\cdot 10^{\textrm{-2}}&&\textrm{, \phantom{-}2.694089739963}\cdot 10^{\textrm{-2}}&&\textrm{, -3.016094127313}\cdot 10^{\textrm{-1}}                  &&\textrm{)}\\
    p_6            &=\textrm{( \phantom{-}8.744382631133}\cdot 10^{\textrm{-2}}&&\textrm{, -1.439660680884}\cdot 10^{\textrm{-1}}&&\textrm{, -2.900615690572}\cdot 10^{\textrm{-1}}&&\textrm{, -1.677299655434}\cdot 10^{\textrm{\phantom{-}0}}                  &&\textrm{)}\\
    p_7            &=\textrm{( -3.863445525253}\cdot 10^{\textrm{-2}}&&\textrm{, \phantom{-}2.503037814880}\cdot 10^{\textrm{-2}}&&\textrm{, -1.989872023809}\cdot 10^{\textrm{-3}}&&\textrm{, -2.317380284719}\cdot 10^{\textrm{-1}}                  &&\textrm{)}\\
    p_8            &=\textrm{( -1.638609443119}\cdot 10^{\textrm{-1}}&&\textrm{, \phantom{-}2.388016861962}\cdot 10^{\textrm{-2}}&&\textrm{, \phantom{-}4.606322903831}\cdot 10^{\textrm{-2}}&&\textrm{, -8.597600332874}\cdot 10^{\textrm{-1}}                  &&\textrm{)}\\
    p_9            &=\textrm{( \phantom{-}6.339609085730}\cdot 10^{\textrm{-2}}&&\textrm{, -5.151857367263}\cdot 10^{\textrm{-2}}&&\textrm{, \phantom{-}7.732066709885}\cdot 10^{\textrm{-2}}&&\textrm{, -5.629545569995}\cdot 10^{\textrm{-1}}                  &&\textrm{)}\\
    p_{10}            &=\textrm{( \phantom{-}1.619398729339}\cdot 10^{\textrm{-2}}&&\textrm{, \phantom{-}3.826976752060}\cdot 10^{\textrm{-2}}&&\textrm{, \phantom{-}7.442114811470}\cdot 10^{\textrm{-2}}&&\textrm{, -4.269170767623}\cdot 10^{\textrm{-1}}                  &&\textrm{)}
  \end{align*}
\end{footnotesize}

For this decagon topology, the masses $m_i$ of the loop propagators denoted $k_i$ in the above figure are set to be equal to $0.1\times i$ (\emph{i.e.} their successive values range from $0.1$ to $1.0$).

\newpage
Kinematics for the triacontagon b):

\begin{footnotesize}
\begin{align*}
    [GeV]         &\phantom{=\textrm{( 0}}E&&           \phantom{\textrm{,0}}p_x                            &&\phantom{\textrm{,0}}p_y                           &&\phantom{\textrm{,0}}p_z                           &&\\
    p_1            &=\textrm{( \phantom{-}4.497499305169}\cdot 10^{\textrm{-2}}&&                           \textrm{, \phantom{-}0}                               &&\textrm{, \phantom{-}0}                               &&\textrm{, \phantom{-}2.5}                           &&\textrm{)}\\    
    p_2            &=\textrm{( -4.497499305169}\cdot 10^{\textrm{-2}}&&                           \textrm{, \phantom{-}0}                               &&\textrm{, \phantom{-}0}                               &&\textrm{, \phantom{-}2.5}                                            &&\textrm{)}\\
    p_{3}            &=\textrm{( \phantom{-}1.190914841688}\cdot 10^{\textrm{-3}}&&\textrm{, \phantom{-}4.811227902221}\cdot 10^{\textrm{-4}}&&\textrm{, -8.341860844659}\cdot 10^{\textrm{-4}}&&\textrm{, -1.071870295668}\cdot 10^{\textrm{-1}}                  &&\textrm{)}\\
    p_{4}            &=\textrm{( -2.953676144156}\cdot 10^{\textrm{-4}}&&\textrm{, \phantom{-}1.653208835624}\cdot 10^{\textrm{-3}}&&\textrm{, \phantom{-}2.848734934414}\cdot 10^{\textrm{-3}}&&\textrm{, -1.815601292161}\cdot 10^{\textrm{-1}}                  &&\textrm{)}\\
    p_{5}            &=\textrm{( -3.076524095310}\cdot 10^{\textrm{-4}}&&\textrm{, \phantom{-}1.513401940618}\cdot 10^{\textrm{-3}}&&\textrm{, \phantom{-}7.861188448553}\cdot 10^{\textrm{-4}}&&\textrm{, -1.145973197461}\cdot 10^{\textrm{-1}}                  &&\textrm{)}\\
    p_{6}            &=\textrm{( \phantom{-}1.826390036566}\cdot 10^{\textrm{-3}}&&\textrm{, -1.970736165619}\cdot 10^{\textrm{-3}}&&\textrm{, -6.152450853423}\cdot 10^{\textrm{-3}}&&\textrm{, -3.439554953186}\cdot 10^{\textrm{-1}}                  &&\textrm{)}\\
    p_{7}            &=\textrm{( -9.504372484440}\cdot 10^{\textrm{-4}}&&\textrm{, \phantom{-}7.908062483944}\cdot 10^{\textrm{-4}}&&\textrm{, \phantom{-}6.339866998243}\cdot 10^{\textrm{-5}}&&\textrm{, -9.724614230247}\cdot 10^{\textrm{-2}}                  &&\textrm{)}\\
    p_{8}            &=\textrm{( -3.995306507647}\cdot 10^{\textrm{-3}}&&\textrm{, \phantom{-}1.291832979641}\cdot 10^{\textrm{-3}}&&\textrm{, \phantom{-}1.465796670893}\cdot 10^{\textrm{-3}}&&\textrm{, -2.346799225421}\cdot 10^{\textrm{-1}}                  &&\textrm{)}\\
    p_{9}            &=\textrm{( \phantom{-}1.397668611845}\cdot 10^{\textrm{-3}}&&\textrm{, -7.382426010659}\cdot 10^{\textrm{-4}}&&\textrm{, \phantom{-}2.042622601469}\cdot 10^{\textrm{-3}}&&\textrm{, -1.493384309095}\cdot 10^{\textrm{-1}}                  &&\textrm{)}\\
    p_{10}            &=\textrm{( \phantom{-}2.922013023842}\cdot 10^{\textrm{-4}}&&\textrm{, \phantom{-}1.278514318475}\cdot 10^{\textrm{-3}}&&\textrm{, \phantom{-}1.951856314299}\cdot 10^{\textrm{-3}}&&\textrm{, -1.394607788531}\cdot 10^{\textrm{-1}}                  &&\textrm{)}\\
    p_{11}            &=\textrm{( \phantom{-}2.083295515367}\cdot 10^{\textrm{-3}}&&\textrm{, -1.206148100689}\cdot 10^{\textrm{-3}}&&\textrm{, \phantom{-}1.196453807047}\cdot 10^{\textrm{-3}}&&\textrm{, -1.539189328635}\cdot 10^{\textrm{-1}}                  &&\textrm{)}\\
    p_{12}            &=\textrm{( \phantom{-}1.119494688534}\cdot 10^{\textrm{-3}}&&\textrm{, \phantom{-}4.323911541453}\cdot 10^{\textrm{-4}}&&\textrm{, \phantom{-}9.659621201751}\cdot 10^{\textrm{-5}}&&\textrm{, -9.617121835581}\cdot 10^{\textrm{-2}}                  &&\textrm{)}\\
    p_{13}            &=\textrm{( \phantom{-}5.716287720297}\cdot 10^{\textrm{-4}}&&\textrm{, \phantom{-}1.106557732182}\cdot 10^{\textrm{-3}}&&\textrm{, -3.298377382829}\cdot 10^{\textrm{-3}}&&\textrm{, -1.915758519962}\cdot 10^{\textrm{-1}}                  &&\textrm{)}\\
    p_{14}            &=\textrm{( -5.369562427291}\cdot 10^{\textrm{-4}}&&\textrm{, -8.116977264731}\cdot 10^{\textrm{-4}}&&\textrm{, \phantom{-}1.165446985452}\cdot 10^{\textrm{-3}}&&\textrm{, -1.067174058163}\cdot 10^{\textrm{-1}}                  &&\textrm{)}\\
    p_{15}            &=\textrm{( -3.038203642400}\cdot 10^{\textrm{-4}}&&\textrm{, \phantom{-}9.971571409418}\cdot 10^{\textrm{-4}}&&\textrm{, \phantom{-}2.249169047901}\cdot 10^{\textrm{-3}}&&\textrm{, -1.448740020112}\cdot 10^{\textrm{-1}}                  &&\textrm{)}\\
    p_{16}            &=\textrm{( \phantom{-}3.487267645290}\cdot 10^{\textrm{-5}}&&\textrm{, \phantom{-}1.615148442435}\cdot 10^{\textrm{-3}}&&\textrm{, -8.542986948583}\cdot 10^{\textrm{-4}}&&\textrm{, -1.182132274187}\cdot 10^{\textrm{-1}}                  &&\textrm{)}\\
    p_{17}            &=\textrm{( -5.117047914552}\cdot 10^{\textrm{-4}}&&\textrm{, \phantom{-}2.247503633421}\cdot 10^{\textrm{-4}}&&\textrm{, -1.651133441512}\cdot 10^{\textrm{-3}}&&\textrm{, -1.149847398690}\cdot 10^{\textrm{-1}}                  &&\textrm{)}\\
    p_{18}            &=\textrm{( -2.846706995927}\cdot 10^{\textrm{-3}}&&\textrm{, -1.362233500134}\cdot 10^{\textrm{-3}}&&\textrm{, \phantom{-}4.794467816646}\cdot 10^{\textrm{-3}}&&\textrm{, -2.966325271881}\cdot 10^{\textrm{-1}}                  &&\textrm{)}\\
    p_{19}            &=\textrm{( \phantom{-}6.091584485757}\cdot 10^{\textrm{-3}}&&\textrm{, \phantom{-}2.195915724549}\cdot 10^{\textrm{-3}}&&\textrm{, -1.197142659379}\cdot 10^{\textrm{-3}}&&\textrm{, -3.376854970358}\cdot 10^{\textrm{-1}}                  &&\textrm{)}\\
    p_{20}            &=\textrm{( \phantom{-}5.980168657704}\cdot 10^{\textrm{-4}}&&\textrm{, \phantom{-}2.629583881849}\cdot 10^{\textrm{-5}}&&\textrm{, -7.665766571385}\cdot 10^{\textrm{-4}}&&\textrm{, -8.938617642223}\cdot 10^{\textrm{-2}}                  &&\textrm{)}\\
    p_{21}            &=\textrm{( -3.342042485085}\cdot 10^{\textrm{-3}}&&\textrm{, \phantom{-}1.650655929455}\cdot 10^{\textrm{-3}}&&\textrm{, \phantom{-}1.312973979301}\cdot 10^{\textrm{-3}}&&\textrm{, -2.113516834560}\cdot 10^{\textrm{-1}}                  &&\textrm{)}\\
    p_{22}            &=\textrm{( -1.264861426574}\cdot 10^{\textrm{-4}}&&\textrm{, \phantom{-}1.915890345560}\cdot 10^{\textrm{-4}}&&\textrm{, \phantom{-}1.316019919940}\cdot 10^{\textrm{-3}}&&\textrm{, -1.004317371667}\cdot 10^{\textrm{-1}}                  &&\textrm{)}\\
    p_{23}            &=\textrm{( \phantom{-}3.562605710800}\cdot 10^{\textrm{-4}}&&\textrm{, -7.763262588944}\cdot 10^{\textrm{-4}}&&\textrm{, \phantom{-}7.852910865420}\cdot 10^{\textrm{-4}}&&\textrm{, -9.481938284651}\cdot 10^{\textrm{-2}}                  &&\textrm{)}\\
    p_{24}            &=\textrm{( -1.421836695497}\cdot 10^{\textrm{-3}}&&\textrm{, -7.456157604865}\cdot 10^{\textrm{-4}}&&\textrm{, -3.770019054652}\cdot 10^{\textrm{-4}}&&\textrm{, -1.114640400272}\cdot 10^{\textrm{-1}}                  &&\textrm{)}\\
    p_{25}            &=\textrm{( -1.310529030416}\cdot 10^{\textrm{-3}}&&\textrm{, -8.542658344383}\cdot 10^{\textrm{-4}}&&\textrm{, -1.598285983984}\cdot 10^{\textrm{-3}}&&\textrm{, -1.346455953796}\cdot 10^{\textrm{-1}}                  &&\textrm{)}\\
    p_{26}            &=\textrm{( -1.913722145629}\cdot 10^{\textrm{-3}}&&\textrm{, \phantom{-}8.006367042237}\cdot 10^{\textrm{-5}}&&\textrm{, -5.783245758276}\cdot 10^{\textrm{-6}}&&\textrm{, -1.216426743368}\cdot 10^{\textrm{-1}}                  &&\textrm{)}\\
    p_{27}            &=\textrm{( \phantom{-}2.334791001581}\cdot 10^{\textrm{-3}}&&\textrm{, \phantom{-}5.082287378374}\cdot 10^{\textrm{-4}}&&\textrm{, -3.533908119629}\cdot 10^{\textrm{-4}}&&\textrm{, -1.421656651323}\cdot 10^{\textrm{-1}}                  &&\textrm{)}\\
    p_{28}            &=\textrm{( -9.648375412153}\cdot 10^{\textrm{-5}}&&\textrm{, \phantom{-}5.048990265442}\cdot 10^{\textrm{-5}}&&\textrm{, -5.142668558715}\cdot 10^{\textrm{-4}}&&\textrm{, -7.947214518508}\cdot 10^{\textrm{-2}}                  &&\textrm{)}\\
    p_{29}            &=\textrm{( \phantom{-}3.717510139732}\cdot 10^{\textrm{-5}}&&\textrm{, -4.957556091852}\cdot 10^{\textrm{-3}}&&\textrm{, -4.157498399677}\cdot 10^{\textrm{-3}}&&\textrm{, -3.320901132800}\cdot 10^{\textrm{-1}}                  &&\textrm{)}\\
    p_{30}            &=\textrm{( \phantom{-}2.475795734193}\cdot 10^{\textrm{-5}}&&\textrm{, -2.665308744662}\cdot 10^{\textrm{-3}}&&\textrm{, -3.145539144343}\cdot 10^{\textrm{-4}}&&\textrm{, -1.537321357582}\cdot 10^{\textrm{-1}}                  &&\textrm{)}
  \end{align*}
\end{footnotesize}

All internal lines are set massless in the triacontagon case (as denoted by the absence of a star next to its label b)).

\newpage
\subsection{Topology c) and c)*: two-loop six-point diagram}

\begin{center}
\begin{tikzpicture}
\pgfmathsetmacro{\r}{1.5}
\pgfmathsetmacro{\n}{6}
\pgfmathsetmacro{\angle}{360/\n}
\foreach \y in {1,...,\n}{
	\pgfmathsetmacro{\start}{(\y+0.5)*\angle}
	\pgfmathsetmacro{\stop}{(\y+1.5)*\angle}
	\pgfmathsetmacro{\mid}{(\y)*\angle}
	\draw [domain=\start:\stop] plot ({\r*cos(\x)}, {\r*sin(\x)});
}
\draw[] (0,\r) --(0,\r/3);
\draw[] (0,\r/3) --(0,-\r/3);
\draw[] (0,-\r/3) --(0,-\r);

\draw[->-] (\r/2,\r/3) node[right]{$p_5$} -- (0,\r/3);
\draw[->-] (\r/2,-\r/3) node[right]{$p_6$} -- (0,-\r/3) ;

\pgfmathsetmacro{\start}{(2+0.5)*\angle}
\draw [->-] (-\r-0.5,{\r*sin(\start)}) node[left]{$p_1$} -- ({\r*cos(\start)}, {\r*sin(\start)});
\pgfmathsetmacro{\start}{(3+0.5)*\angle}
\draw [->-](-\r-0.5,{\r*sin(\start)}) node[left]{$p_2$} -- ({\r*cos(\start)}, {\r*sin(\start)});
\pgfmathsetmacro{\start}{(5+0.5)*\angle}
\draw [->-]  (\r+0.5,{\r*sin(\start)}) node[right]{$p_3$} -- ({\r*cos(\start)}, {\r*sin(\start)});
\pgfmathsetmacro{\start}{(6+0.5)*\angle}
\draw [->-] (\r+0.5,{\r*sin(\start)}) node[right]{$p_4$} --({\r*cos(\start)}, {\r*sin(\start)});

\end{tikzpicture}
\end{center}

\begin{footnotesize}
\begin{align*}
    [GeV]         &\phantom{=\textrm{( 0}}E&&           \phantom{\textrm{,0}}p_x                            &&\phantom{\textrm{,0}}p_y                           &&\phantom{\textrm{,0}}p_z                           &&\\
    p_{1}            &=\textrm{( \phantom{-}2.000000000000}\cdot 10^{\textrm{-1}}&&\textrm{, \phantom{-}3.000000000000}\cdot 10^{\textrm{-1}}&&\textrm{, \phantom{-}5.000000000000}\cdot 10^{\textrm{-1}}&&\textrm{, \phantom{-}6.000000000000}\cdot 10^{\textrm{-1}}                  &&\textrm{)}\\
    p_{2}            &=\textrm{( -1.000000000000}\cdot 10^{\textrm{-1}}&&\textrm{, \phantom{-}7.000000000000}\cdot 10^{\textrm{-1}}&&\textrm{, \phantom{-}2.000000000000}\cdot 10^{\textrm{-1}}&&\textrm{, \phantom{-}1.000000000000}\cdot 10^{\textrm{-1}}                  &&\textrm{)}\\
    p_{3}            &=\textrm{( \phantom{-}1.000000000000}\cdot 10^{\textrm{-1}}&&\textrm{, \phantom{-}5.000000000000}\cdot 10^{\textrm{-1}}&&\textrm{, -3.000000000000}\cdot 10^{\textrm{-1}}&&\textrm{, -4.000000000000}\cdot 10^{\textrm{-1}}                  &&\textrm{)}\\
    p_{4}            &=\textrm{( -3.000000000000}\cdot 10^{\textrm{-1}}&&\textrm{, \phantom{-}4.000000000000}\cdot 10^{\textrm{-1}}&&\textrm{, \phantom{-}5.000000000000}\cdot 10^{\textrm{-1}}&&\textrm{, \phantom{-}2.000000000000}\cdot 10^{\textrm{-1}}                  &&\textrm{)}\\
    p_{5}            &=\textrm{( -2.000000000000}\cdot 10^{\textrm{-1}}&&\textrm{, \phantom{-}3.000000000000}\cdot 10^{\textrm{-1}}&&\textrm{, \phantom{-}2.000000000000}\cdot 10^{\textrm{-1}}&&\textrm{, -5.000000000000}\cdot 10^{\textrm{-1}}                  &&\textrm{)}\\
    p_{6}            &=\textrm{( \phantom{-}3.000000000000}\cdot 10^{\textrm{-1}}&&\textrm{, -2.200000000000}\cdot 10^{\textrm{\phantom{-}0}}&&\textrm{, -1.100000000000}\cdot 10^{\textrm{\phantom{-}0}}&&\textrm{, \phantom{-}0.000000000000}\cdot 10^{\textrm{\phantom{-}0}}                  &&\textrm{)}\\
    \end{align*}
\end{footnotesize}

The mass of each loop propagator is set to $1$ in the massive c)* case reported in Table I of the main text and massless in the c) case.

\subsection{Topology d)*: two-loop eight-point diagram}
\begin{center}
\begin{tikzpicture}
	\pgfmathsetmacro{\r}{1.5}
	\pgfmathsetmacro{\n}{6}
	\pgfmathsetmacro{\angle}{360/\n}
	\foreach \y in {1,...,\n}{
		\pgfmathsetmacro{\start}{(\y+0.5)*\angle}
		\pgfmathsetmacro{\stop}{(\y+1.5)*\angle}
		\pgfmathsetmacro{\mid}{(\y)*\angle}
		\draw [domain=\start:\stop] plot ({\r*cos(\x)}, {\r*sin(\x)});
	}
	\draw[] (0,\r) --(0,\r/5);
	\draw[] (0,\r/5) --(0,-\r/5);
	\draw[] (0,-\r/5) --(0,-\r);
	
	\draw[->-] (\r/2,\r/5) node[right]{$p_6$} -- (0,\r/5);
	\draw[->-] (\r/2,\r/5*2) node[right]{$p_5$} -- (0,\r/5*2);
	\draw[->-] (\r/2,-\r/5) node[right]{$p_7$} -- (0,-\r/5);
	\draw[->-] (\r/2,-\r/5*2) node[right]{$p_8$} -- (0,-\r/5*2);
	
	\pgfmathsetmacro{\start}{(2+0.5)*\angle}
	\draw [->-] (-\r-0.5,{\r*sin(\start)}) node[left]{$p_1$} -- ({\r*cos(\start)}, {\r*sin(\start)});
	\pgfmathsetmacro{\start}{(3+0.5)*\angle}
	\draw [->-](-\r-0.5,{\r*sin(\start)}) node[left]{$p_2$} -- ({\r*cos(\start)}, {\r*sin(\start)});
	\pgfmathsetmacro{\start}{(5+0.5)*\angle}
	\draw [->-]  (\r+0.5,{\r*sin(\start)}) node[right]{$p_3$} -- ({\r*cos(\start)}, {\r*sin(\start)});
	\pgfmathsetmacro{\start}{(6+0.5)*\angle}
	\draw [->-] (\r+0.5,{\r*sin(\start)}) node[right]{$p_4$} --({\r*cos(\start)}, {\r*sin(\start)});
	
\end{tikzpicture}
\end{center}

\begin{footnotesize}
\begin{align*}
    [GeV]         &\phantom{=\textrm{( 0}}E&&           \phantom{\textrm{,0}}p_x                            &&\phantom{\textrm{,0}}p_y                           &&\phantom{\textrm{,0}}p_z                           &&\\
    p_{1}            &=\textrm{( \phantom{-}1.500000000000}\cdot 10^{\textrm{-1}}&&\textrm{, \phantom{-}9.000000000000}\cdot 10^{\textrm{-2}}&&\textrm{, \phantom{-}2.300000000000}\cdot 10^{\textrm{-1}}&&\textrm{, \phantom{-}0.000000000000}\cdot 10^{\textrm{\phantom{-}0}}                  &&\textrm{)}\\
    p_{2}            &=\textrm{( -2.000000000000}\cdot 10^{\textrm{-1}}&&\textrm{, \phantom{-}7.900000000000}\cdot 10^{\textrm{-1}}&&\textrm{, \phantom{-}1.200000000000}\cdot 10^{\textrm{-1}}&&\textrm{, \phantom{-}1.100000000000}\cdot 10^{\textrm{-1}}                  &&\textrm{)}\\
    p_{3}            &=\textrm{( -2.300000000000}\cdot 10^{\textrm{-1}}&&\textrm{, \phantom{-}1.400000000000}\cdot 10^{\textrm{-1}}&&\textrm{, -4.700000000000}\cdot 10^{\textrm{-1}}&&\textrm{, -2.200000000000}\cdot 10^{\textrm{-1}}                  &&\textrm{)}\\
    p_{4}            &=\textrm{( \phantom{-}1.100000000000}\cdot 10^{\textrm{-1}}&&\textrm{, -5.900000000000}\cdot 10^{\textrm{-1}}&&\textrm{, \phantom{-}5.400000000000}\cdot 10^{\textrm{-1}}&&\textrm{, \phantom{-}1.200000000000}\cdot 10^{\textrm{-1}}                  &&\textrm{)}\\
    p_{5}            &=\textrm{( -1.500000000000}\cdot 10^{\textrm{-1}}&&\textrm{, \phantom{-}2.100000000000}\cdot 10^{\textrm{-1}}&&\textrm{, \phantom{-}1.000000000000}\cdot 10^{\textrm{-1}}&&\textrm{, -3.200000000000}\cdot 10^{\textrm{-1}}                  &&\textrm{)}\\
    p_{6}            &=\textrm{( \phantom{-}3.200000000000}\cdot 10^{\textrm{-1}}&&\textrm{, \phantom{-}8.400000000000}\cdot 10^{\textrm{-1}}&&\textrm{, \phantom{-}2.700000000000}\cdot 10^{\textrm{-1}}&&\textrm{, \phantom{-}4.900000000000}\cdot 10^{\textrm{-1}}                  &&\textrm{)}\\
    p_{7}            &=\textrm{( \phantom{-}1.100000000000}\cdot 10^{\textrm{-1}}&&\textrm{, -3.000000000000}\cdot 10^{\textrm{-1}}&&\textrm{, -1.200000000000}\cdot 10^{\textrm{-1}}&&\textrm{, -1.000000000000}\cdot 10^{\textrm{-1}}                  &&\textrm{)}\\
    p_{8}            &=\textrm{( -1.100000000000}\cdot 10^{\textrm{-1}}&&\textrm{, -1.180000000000}\cdot 10^{\textrm{\phantom{-}0}}&&\textrm{, -6.700000000000}\cdot 10^{\textrm{-1}}&&\textrm{, -8.000000000000}\cdot 10^{\textrm{-2}}                  &&\textrm{)}\\
    \end{align*}
\end{footnotesize}

Each internal line has a mass of $1$.

\subsection{Topology e): four-loop four-point ladder diagram}

\begin{center}
\begin{tikzpicture}
	\pgfmathsetmacro{\r}{0.75}
	\pgfmathsetmacro{\rtwo}{1.5*\r}
	\pgfmathsetmacro{\rfour}{4*\r}
	\pgfmathsetmacro{\rleg}{\r}
	\begin{feynman}
	
	\tikzfeynmanset{every vertex={empty dot,minimum size=0mm}}
	 \vertex (a1);
	\vertex[right=\r of a1] (a2);
	\vertex[above=\rtwo of a2] (a4);
	\vertex[below=\rtwo of a2] (a5);

	\vertex[right=\rfour of a1] (d1);

	\vertex[right=\r of a4] (b1);
	\vertex[right=\r of a5] (b2);
	\vertex[right=\r of b1] (c1);
	\vertex[right=\r of b2] (c2);
	
	\vertex[above=\rtwo of a1] (a7);
	\vertex[below=\rtwo of a1] (a8);
	\vertex[above=\rtwo of d1] (a9);
	\vertex[below=\rtwo of d1] (a10);
	
	\vertex[left=\rleg of a7] (a11);
	\vertex[left=\rleg of a8] (a12);
	\vertex[right=\rleg of a9] (a13);
	\vertex[right=\rleg of a10] (a14);
	

	\vertex[above=\rtwo of a1] (e1);
	\vertex[below=\rtwo of a1] (e2);    
	
	\vertex[above=\rtwo of d1] (f1);
	\vertex[below=\rtwo of d1] (f2);            

		\diagram*{	
		(e1)--(a4), 
		(a5)--(e2),
(e1)--(e2),

		(a4) -- (a5),
		(a4) -- (b1),
		(a5) -- (b2),
		(b1)--(b2),
		(b1)--(c1),
		(b2)--(c2),
		(c1)--(c2),

		(c1) --  (f1),
		(f2) -- (c2),
		(f1)--(f2),
		
		(a7) -- (a11),
		(a8) -- (a12),
		(a9) -- (a13),
		(a10) -- (a14),
		};

		\draw [->-] (a11) node[left]{$p_1$} --(a7);
		\draw [->-] (a12) node[left]{$p_2$} --(a8);
		\draw [->-] (a13) node[right]{$p_4$} --(a9);
		\draw [->-] (a14) node[right]{$p_3$} --(a10);
	\end{feynman}
\end{tikzpicture}
\end{center}

\begin{footnotesize}
\begin{align*}
    [GeV]         &\phantom{=\textrm{( 0}}E&&           \phantom{\textrm{,0}}p_x                            &&\phantom{\textrm{,0}}p_y                           &&\phantom{\textrm{,0}}p_z                           &&\\
     p_{1}            &=\textrm{( \phantom{-}1.000000000000}\cdot 10^{\textrm{-1}}&&\textrm{, \phantom{-}2.000000000000}\cdot 10^{\textrm{-1}}&&\textrm{, \phantom{-}5.000000000000}\cdot 10^{\textrm{-1}}&&\textrm{, \phantom{-}1.000000000000}\cdot 10^{\textrm{-1}}                  &&\textrm{)}\\
    p_{2}            &=\textrm{( -3.000000000000}\cdot 10^{\textrm{-1}}&&\textrm{, \phantom{-}4.000000000000}\cdot 10^{\textrm{-1}}&&\textrm{, \phantom{-}1.000000000000}\cdot 10^{\textrm{-1}}&&\textrm{, \phantom{-}2.000000000000}\cdot 10^{\textrm{-1}}                  &&\textrm{)}\\
    p_{3}            &=\textrm{( \phantom{-}1.000000000000}\cdot 10^{\textrm{-1}}&&\textrm{, \phantom{-}2.000000000000}\cdot 10^{\textrm{-1}}&&\textrm{, \phantom{-}5.000000000000}\cdot 10^{\textrm{-1}}&&\textrm{, \phantom{-}3.000000000000}\cdot 10^{\textrm{-1}}                  &&\textrm{)}\\
    p_{4}            &=\textrm{( \phantom{-}1.000000000000}\cdot 10^{\textrm{-1}}&&\textrm{, -8.000000000000}\cdot 10^{\textrm{-1}}&&\textrm{, -1.100000000000}\cdot 10^{\textrm{\phantom{-}0}}&&\textrm{, -6.000000000000}\cdot 10^{\textrm{-1}}                  &&\textrm{)}\\
 \end{align*}
\end{footnotesize}

\subsection{Topology f): three-loop two-point Mercedes diagram}
\begin{center}
    \begin{tikzpicture}
    
    \pgfmathsetmacro{\r}{2.5}
    \pgfmathsetmacro{\rh}{0.5*\r}
    \pgfmathsetmacro{\rq}{0.25*\r}
    \pgfmathsetmacro{\rqq}{0.15*\r}
	\pgfmathsetmacro{\ra}{0.35355*\r}
	\pgfmathsetmacro{\rb}{0.17678*\r}
	\pgfmathsetmacro{\rc}{0.14645*\r}
    \begin{feynman}
    
    \tikzfeynmanset{every vertex={empty dot,minimum size=0mm}}
    
    \vertex (a1);
    
    \vertex[below=\rh of a1] (a3);
    \vertex[below=\rh of a3] (b1);
    \vertex[below=\ra of a3] (a5);
    \vertex[left=\ra of a5] (a6);
    
    \vertex[below=\rb of a3] (b2);
    
    \vertex[below=\rc of a1] (b3);
    
    \vertex[below=\ra of a3] (b4);
    
    \vertex[left=\rh of a3] (a7);
    \vertex[left=\rqq of a7] (a8);
    
    \vertex[right=\rh of a3] (a9);
    \vertex[right=\rqq of a9] (a10);
    
    \vertex[left=\ra of b3] (a11);
    \vertex[left=\rqq of a11] (a12);
    
    \vertex[below=\rq of a1] (a13);
    \vertex[right=\rqq of a13] (a14);
    
    \vertex[right=\ra of b4] (a15);
    \vertex[right=\rqq of a15] (a16);
    
    \vertex[left=\rb of b2] (a17);
    \vertex[right=\rqq of a17] (a18);

    \vertex[left=\rh of a3] (a2);
    \vertex[right=\rh of a3] (a4);
    
    \vertex[left=\rb of b2] (c1);
    \vertex[below=\rq of a1] (c2);
    
    \vertex[left=\ra of b3] (c3);
    \vertex[right=\ra of b4] (c4);
    
        \diagram*{	
        (a1)--[quarter left](a4) -- [quarter left](b1) -- [quarter left](a2)-- [quarter left](a1),
        (a3) -- (a1),
        (a3) -- (a6),
        (a3) -- (a4),
        }; 
        
        \draw [->-] (a8) node[left]{$p_1$} --(a7);
        \draw [->-] (a10) node[right]{$-p_1$} --(a9);
    \end{feynman}
    \end{tikzpicture}
    \end{center}

\begin{footnotesize}
\begin{align*}
    [GeV]         &\phantom{=\textrm{( 0}}E&&           \phantom{\textrm{,0}}p_x                            &&\phantom{\textrm{,0}}p_y                           &&\phantom{\textrm{,0}}p_z                           &&\\
    p_{1}            &=\textrm{( \phantom{-}0.000000000000}\cdot 10^{\textrm{\phantom{-}0}}&&\textrm{, \phantom{-}0.000000000000}\cdot 10^{\textrm{\phantom{-}0}}&&\textrm{, \phantom{-}0.000000000000}\cdot 10^{\textrm{\phantom{-}0}}&&\textrm{, \phantom{-}1.000000000000}\cdot 10^{\textrm{\phantom{-}0}}                  &&\textrm{)}\\
 \end{align*}
\end{footnotesize}

\subsection{Topology g)*: three-loop six-point Mercedes diagram}

\begin{center}
    \begin{tikzpicture}
    
    \pgfmathsetmacro{\r}{2.5}
    \pgfmathsetmacro{\rh}{0.5*\r}
    \pgfmathsetmacro{\rq}{0.25*\r}
    \pgfmathsetmacro{\rqq}{0.25*\r}
	\pgfmathsetmacro{\ra}{0.35355*\r}
	\pgfmathsetmacro{\rb}{0.17678*\r}
	\pgfmathsetmacro{\rc}{0.14645*\r}
    \begin{feynman}
    
    \tikzfeynmanset{every vertex={empty dot,minimum size=0mm}}
    
    \vertex (a1);
    
    \vertex[below=\rh of a1] (a3);
    \vertex[below=\rh of a3] (b1);
    \vertex[below=\ra of a3] (a5);
    \vertex[left=\ra of a5] (a6);
    
    \vertex[below=\rb of a3] (b2);
    
    \vertex[below=\rc of a1] (b3);
    
    \vertex[below=\ra of a3] (b4);
    
    \vertex[left=\rh of a3] (a7);
    \vertex[left=\rqq of a7] (a8);
    
    \vertex[right=\rh of a3] (a9);
    \vertex[right=\rqq of a9] (a10);
    
    \vertex[left=\ra of b3] (a11);
    \vertex[left=\rqq of a11] (a12);
    
    \vertex[below=\rq of a1] (a13);
    \vertex[right=\rqq of a13] (a14);
    
    \vertex[right=\ra of b4] (a15);
    \vertex[right=\rqq of a15] (a16);
    
    \vertex[left=\rb of b2] (a17);
    \vertex[right=\rqq of a17] (a18);

    \vertex[left=\rh of a3] (a2);
    \vertex[right=\rh of a3] (a4);
    
    \vertex[left=\rb of b2] (c1);
    \vertex[below=\rq of a1] (c2);
    
    \vertex[left=\ra of b3] (c3);
    \vertex[right=\ra of b4] (c4);
    
        \diagram*{	
        (a1)--[quarter left](a4) -- [quarter left](b1) -- [quarter left](a2)-- [quarter left](a1),
        (a3) -- (a1),
        (a3) -- (a6),
        (a3) -- (a4),
        }; 
        
        \draw [->-] (a8) node[left]{$p_1$} --(a7);
        \draw [->-] (a10) node[right]{$p_2$} --(a9);
        \draw [->-] (a12) node[left]{$p_6$} --(a11);
		\draw [->-] (a14) node[right]{$p_5$} --(a13);
        \draw [->-] (a16) node[right]{$p_3$} --(a15);
		\draw [->-] (a18) node[right]{$p_4$} --(a17);
    \end{feynman}
    \end{tikzpicture}
    \end{center}

\begin{footnotesize}
\begin{align*}
    [GeV]         &\phantom{=\textrm{( 0}}E&&           \phantom{\textrm{,0}}p_x                            &&\phantom{\textrm{,0}}p_y                           &&\phantom{\textrm{,0}}p_z                           &&\\
    p_{1}            &=\textrm{( \phantom{-}2.000000000000}\cdot 10^{\textrm{-1}}&&\textrm{, \phantom{-}3.000000000000}\cdot 10^{\textrm{-1}}&&\textrm{, \phantom{-}5.000000000000}\cdot 10^{\textrm{-1}}&&\textrm{, \phantom{-}6.000000000000}\cdot 10^{\textrm{-1}}                  &&\textrm{)}\\
    p_{2}            &=\textrm{( -1.000000000000}\cdot 10^{\textrm{-1}}&&\textrm{, \phantom{-}7.000000000000}\cdot 10^{\textrm{-1}}&&\textrm{, \phantom{-}2.000000000000}\cdot 10^{\textrm{-1}}&&\textrm{, \phantom{-}1.000000000000}\cdot 10^{\textrm{-1}}                  &&\textrm{)}\\
    p_{3}            &=\textrm{( \phantom{-}1.000000000000}\cdot 10^{\textrm{-1}}&&\textrm{, \phantom{-}5.000000000000}\cdot 10^{\textrm{-1}}&&\textrm{, -3.000000000000}\cdot 10^{\textrm{-1}}&&\textrm{, -4.000000000000}\cdot 10^{\textrm{-1}}                  &&\textrm{)}\\
    p_{4}            &=\textrm{( -3.000000000000}\cdot 10^{\textrm{-1}}&&\textrm{, \phantom{-}4.000000000000}\cdot 10^{\textrm{-1}}&&\textrm{, \phantom{-}5.000000000000}\cdot 10^{\textrm{-1}}&&\textrm{, \phantom{-}2.000000000000}\cdot 10^{\textrm{-1}}                  &&\textrm{)}\\
    p_{5}            &=\textrm{( -2.000000000000}\cdot 10^{\textrm{-1}}&&\textrm{, \phantom{-}3.000000000000}\cdot 10^{\textrm{-1}}&&\textrm{, \phantom{-}2.000000000000}\cdot 10^{\textrm{-1}}&&\textrm{, -5.000000000000}\cdot 10^{\textrm{-1}}                  &&\textrm{)}\\
    p_{6}            &=\textrm{( \phantom{-}3.000000000000}\cdot 10^{\textrm{-1}}&&\textrm{, -2.200000000000}\cdot 10^{\textrm{\phantom{-}0}}&&\textrm{, -1.100000000000}\cdot 10^{\textrm{\phantom{-}0}}&&\textrm{, \phantom{-}0.000000000000}\cdot 10^{\textrm{\phantom{-}0}}                  &&\textrm{)}\\
    \end{align*}
\end{footnotesize}

Every internal line has a mass of $1$.

\subsection{Topology h): four-loop two-point non-planar diagram}
 \begin{center}
    \begin{tikzpicture}
        
    \pgfmathsetmacro{\r}{2.5}
    \pgfmathsetmacro{\rhh}{0.25*\r}
    \pgfmathsetmacro{\rh}{0.5*\r}
    \pgfmathsetmacro{\rq}{0.35*\r}
    \pgfmathsetmacro{\rqq}{0.08*\r}
	\pgfmathsetmacro{\ra}{0.35355*\r}
    \begin{feynman}

    \tikzfeynmanset{every vertex={empty dot,minimum size=0mm}}
    
    \vertex (a1);
    
    \vertex[below=\rh of a1] (a3);
    
    \vertex[right=\rqq of a3] (cross1);
    \vertex[left=\rqq of a3] (cross2);
    
    \vertex[below=\rh of a3] (b1);
    
    \vertex[below=\rhh of a1] (c1);
    \vertex[right=\rq of a3] (c2);
    
    \vertex[below=\ra of a3] (a5);
    
    \vertex[left=\rh of a3] (a2);
    
    \vertex[right=\rh of a3] (a9);
    \vertex[right=\rhh of a9] (a10);
    
    \vertex[left=\ra of a5] (a11);
    \vertex[left=\rhh of a11] (a12);
    
    \vertex[right=\rh of a3] (a4);
    \vertex[left=\ra of a5] (a6);

        \diagram*{	
        (a1)--[quarter left](a4) -- [quarter left](b1) -- [quarter left](a2)-- [quarter left](a1),
        (a1) -- (c1),
        (c1) -- (c2),
        (c2) -- (a4),
        (c2) -- (cross1),
        (a2) -- (cross2),
        (cross1) --[white] (cross2),	
        (c1) -- (b1),
        }; 
    \end{feynman}
        \draw [->-] (a12) node[left]{$p_1$} --(a11);
        \draw [->-] (a10) node[right]{$-p_1$} --(a9);
    \end{tikzpicture}
    \end{center}
    
\begin{footnotesize}
\begin{align*}
    [GeV]         &\phantom{=\textrm{( 0}}E&&           \phantom{\textrm{,0}}p_x                            &&\phantom{\textrm{,0}}p_y                           &&\phantom{\textrm{,0}}p_z                           &&\\
    p_{1}            &=\textrm{( \phantom{-}0.000000000000}\cdot 10^{\textrm{\phantom{-}0}}&&\textrm{, \phantom{-}0.000000000000}\cdot 10^{\textrm{\phantom{-}0}}&&\textrm{, \phantom{-}0.000000000000}\cdot 10^{\textrm{\phantom{-}0}}&&\textrm{, \phantom{-}1.000000000000}\cdot 10^{\textrm{\phantom{-}0}}                  &&\textrm{)}\\
 \end{align*}
\end{footnotesize}

\subsection{Double-box of sect. IV that requires deformation}
\begin{center}
\begin{tikzpicture}
	\pgfmathsetmacro{\r}{0.75}
	\pgfmathsetmacro{\rtwo}{1.5*\r}
	\pgfmathsetmacro{\rend}{2*\r}
	\pgfmathsetmacro{\rleg}{\r}
	\begin{feynman}
	
	\tikzfeynmanset{every vertex={empty dot,minimum size=0mm}}
	 \vertex (a1);
	\vertex[right=\r of a1] (a2);
	\vertex[above=\rtwo of a2] (a4);
	\vertex[below=\rtwo of a2] (a5);

	\vertex[right=\rend of a1] (d1);

	\vertex[right=\r of a4] (b1);
	\vertex[right=\r of a5] (b2);
	\vertex[right=\r of b1] (c1);
	\vertex[right=\r of b2] (c2);
	
	\vertex[above=\rtwo of a1] (a7);
	\vertex[below=\rtwo of a1] (a8);
	\vertex[above=\rtwo of d1] (a9);
	\vertex[below=\rtwo of d1] (a10);
	
	\vertex[left=\rleg of a7] (a11);
	\vertex[left=\rleg of a8] (a12);
	\vertex[right=\rleg of a9] (a13);
	\vertex[right=\rleg of a10] (a14);

	\vertex[above=\rtwo of a1] (e1);
	\vertex[below=\rtwo of a1] (e2);    
	
	\vertex[above=\rtwo of d1] (f1);
	\vertex[below=\rtwo of d1] (f2);            

		\diagram*{	
		(e1)--(a4), 
		(a5)--(e2),
(e1)--(e2),

		(a4) -- (a5),
		(a4) -- (b1),
		(a5) -- (b2),
		(b1)--(b2),
		(b1)--(c1),

		
		(a7) -- (a11),
		(a8) -- (a12),
		(a9) -- (a13),
		(a10) -- (a14),
		};

		\draw [->-] (a11) node[left]{$p_1$} --(a7);
		\draw [->-] (a12) node[left]{$p_2$} --(a8);
		\draw [->-] (a13) node[right]{$p_4$} --(a9);
		\draw [->-] (a14) node[right]{$p_3$} --(a10);
	\end{feynman}
\end{tikzpicture}
\end{center}

\begin{footnotesize}
\begin{align*}
    [GeV]         &\phantom{=\textrm{( 0}}E&&           \phantom{\textrm{,0}}p_x                            &&\phantom{\textrm{,0}}p_y                           &&\phantom{\textrm{,0}}p_z                           &&\\
    p_{1}            &=\textrm{( \phantom{-}1.965860000000}\cdot 10^{\textrm{\phantom{-}1}}&&\textrm{, -7.152520000000}\cdot 10^{\textrm{\phantom{-}0}}&&\textrm{, -2.060160000000}\cdot 10^{\textrm{-1}}&&\textrm{, \phantom{-}8.963830000000}\cdot 10^{\textrm{\phantom{-}0}}                  &&\textrm{)}\\
    p_{2}            &=\textrm{( \phantom{-}2.687400000000}\cdot 10^{\textrm{\phantom{-}1}}&&\textrm{, \phantom{-}7.042030000000}\cdot 10^{\textrm{\phantom{-}0}}&&\textrm{, -5.012950000000}\cdot 10^{\textrm{-2}}&&\textrm{, -1.290550000000}\cdot 10^{\textrm{\phantom{-}1}}                  &&\textrm{)}\\
    p_{3}            &=\textrm{( \phantom{-}4.346740000000}\cdot 10^{\textrm{\phantom{-}1}}&&\textrm{, \phantom{-}1.104910000000}\cdot 10^{\textrm{-1}}&&\textrm{, \phantom{-}2.561460000000}\cdot 10^{\textrm{-1}}&&\textrm{, \phantom{-}3.941700000000}\cdot 10^{\textrm{\phantom{-}0}}                  &&\textrm{)}\\
    p_{4}            &=\textrm{( -9.000000000000}\cdot 10^{\textrm{\phantom{-}1}}&&\textrm{, \phantom{-}0.000000000000}\cdot 10^{\textrm{\phantom{-}0}}&&\textrm{, \phantom{-}0.000000000000}\cdot 10^{\textrm{\phantom{-}0}}&&\textrm{, \phantom{-}0.000000000000}\cdot 10^{\textrm{\phantom{-}0}}                  &&\textrm{)}\\
 \end{align*}
\end{footnotesize}

\end{document}